\documentclass{article}
\usepackage{amsmath,amssymb,amsthm,amsfonts,fullpage,graphics,authblk}

\title{On the modified nonlinear Schr\"odinger equation in the semiclassical limit:  supersonic, subsonic, and transsonic behavior}
\author[a]{Jeffery C. DiFranco}
\author[b]{Peter D. Miller\footnote{The work of the first two authors was partially supported by the National Science Foundation under grant DMS-0807653.}}
\author[b]{Benson K. Muite}
\affil[a]{Department of Mathematics, Seattle University, 
901 12th Avenue,
Seattle, WA 98122}
\affil[b]{Department of Mathematics, University of Michigan, East Hall, 530 Church St., Ann Arbor, MI 48109}
\begin{document}
\maketitle
\begin{abstract}
The purpose of this paper is to present a comparison between the modified nonlinear Schr\"odinger (MNLS) equation and the focusing and defocusing variants of the (unmodified) nonlinear Schr\"odinger (NLS) equation in the semiclassical limit.  We describe aspects of the limiting dynamics and discuss how the nature of the dynamics is evident theoretically through inverse-scattering and noncommutative steepest descent methods.   The main message is that, depending on initial data, the MNLS equation can behave either like the defocusing NLS equation, like the focusing NLS equation (in both cases the analogy is asymptotically accurate in the semiclassical limit when the NLS equation is posed with appropriately modified initial data), or like an interesting mixture of the two.  In the latter case, we identify a feature of the dynamics analogous to a sonic line in gas dynamics, a free boundary separating subsonic flow from supersonic flow.
\end{abstract}
\section{Introduction}
This paper is concerned with the semiclassically scaled modified nonlinear Schr\"odinger (MNLS) equation:
\begin{equation}
i\epsilon\frac{\partial\phi}{\partial t} + \frac{1}{2}\epsilon^2\frac{\partial^2\phi}{\partial x^2} +
|\phi|^2\phi + i\alpha\epsilon\frac{\partial}{\partial x}(|\phi|^2\phi)=0,\quad \alpha,\epsilon>0,
\label{eq:MNLS}
\end{equation}
which can be thought of as a perturbation (with perturbation parameter $\alpha$) of the 
nonlinear Schr\"odinger (NLS) equation:
\begin{equation}
i\epsilon\frac{\partial\phi}{\partial t} +\frac{1}{2}\epsilon^2\frac{\partial^2\phi}{\partial x^2} + 
\kappa |\phi|^2\phi=0,\quad \epsilon>0
\label{eq:NLS}
\end{equation}
taken in the so-called focusing case of
$\kappa=+1$.  (For $\kappa=-1$ one has the so-called defocusing case.)  Our goal is to 
describe connections between the MNLS equation on the one hand and the NLS equation
(perhaps surprisingly, both focusing and defocusing types are related to MNLS) on the other, connections that become especially clear in the semiclassical limit $\epsilon\ll 1$.  

\subsection{Formal considerations.}
\label{sec:formal}
Erwin Madelung was probably the first person to propose thinking of Schr\"odinger-type equations for a complex field $\phi=\phi_\epsilon(x,t)$ as describing the motion of a kind of fluid \cite{Madelung26}.  He introduced the real-valued fluid dynamical variables $\rho_\epsilon$ (density) and $u_\epsilon$ (velocity) given in terms of $\phi_\epsilon$ by
\begin{equation}
\rho_\epsilon(x,t):=|\phi_\epsilon(x,t)|^2\quad\text{and}\quad u_\epsilon(x,t):=\epsilon\Im\left\{\frac{\partial}{\partial x}\log(\phi_\epsilon(x,t))\right\}.
\label{eq:MadelungVars}
\end{equation}
Here $\log(\phi_\epsilon(x,t))$ is defined by continuation to be a smooth function of $x$.
A key observation is that if the complex-valued field $\phi_\epsilon(x,t)$ initially has the form
of an oscillatory wavepacket:
\begin{equation}
\phi_\epsilon(x,0)=A(x)e^{iS(x)/\epsilon}
\label{eq:wavepacket}
\end{equation}
where $A$ and $S$ are real-valued functions independent of $\epsilon$, then the corresponding initial values of the Madelung variables are independent of $\epsilon$ as well:
\begin{equation}
\rho_0(x):=\rho_\epsilon(x,0)=A(x)^2\quad\text{and}\quad u_0(x):= u_\epsilon(x,0)=S'(x).
\label{eq:Madelungwavepacket}
\end{equation}
Without approximation, the NLS equation \eqref{eq:NLS} can be written in terms of the Madelung variables as:
\begin{equation}
\frac{\partial\rho_\epsilon}{\partial t} + \frac{\partial}{\partial x}(\rho_\epsilon u_\epsilon)=0\quad\text{and}\quad
\frac{\partial u_\epsilon}{\partial t} +\frac{\partial}{\partial x}\left(\frac{1}{2}u_\epsilon^2-\kappa\rho_\epsilon\right)=\frac{1}{2}\epsilon^2\frac{\partial F[\rho_\epsilon]}{\partial x}
\label{eq:MadelungNLS}
\end{equation}
where $F[\rho]$ denotes the expression
\begin{equation}
F[\rho]:=\frac{1}{2\rho}\frac{\partial^2\rho}{\partial x^2}-\left(\frac{1}{2\rho}\frac{\partial\rho}{\partial x}\right)^2.
\label{eq:Fdef}
\end{equation}
With $\epsilon$-independent  initial data of the form \eqref{eq:Madelungwavepacket} taken in \eqref{eq:MadelungNLS},
it seems reasonable to neglect the dispersive term $\tfrac{1}{2}\epsilon^2\partial_x F[\rho_\epsilon]$ to obtain limiting $\epsilon$-independent equations that one might expect to govern the leading terms $(\rho(x,t),u(x,t))$ of
the $\epsilon$-dependent fields $(\rho_\epsilon(x,t),u_\epsilon(x,t))$ when $\epsilon\ll 1$.  
The $\epsilon=0$ truncated system of equations 
\begin{equation}
\frac{\partial\rho}{\partial t} +\frac{\partial}{\partial x}(\rho u)=0\quad\text{and}\quad
\frac{\partial u}{\partial t} +\frac{\partial}{\partial x}\left(\frac{1}{2}u^2-\kappa\rho\right)=0,
\label{eq:WhithamNLS}
\end{equation}
which we call the \emph{dispersionless NLS system}, is really a classical Euler system of compressible fluid (gas) dynamics.  The term $-\kappa\rho$ in the flux for $u$ represents the pressure as a function of density.  If $\kappa=-1$ (the defocusing case of the NLS equation) then the pressure is an increasing function of the gas density and the physical dynamics is that of the gas ``trying to get out of its own way''; localized concentrations in density tend to decay with time.  On the other hand, if $\kappa=+1$ (the focusing case of the NLS equation) then the pressure decreases with increasing density and localized concentrations in density are dynamically enhanced.  In physical problems with such pressure laws, like supercooled Van der Waals gases, initially small but nonuniform perturbations of a uniform density lead quickly to spontaneous condensation of the gas into liquid droplets.  Mathematically, the equations \eqref{eq:WhithamNLS} comprise a quasilinear system of hyperbolic type for  $\kappa=-1$, and the Cauchy initial-value problem for such systems is locally well-posed in suitable function spaces.  
On the other hand when $\kappa=+1$ we have instead a system of elliptic type, and the Cauchy
problem is ill-posed.  

The ill-posedness of the Cauchy initial-value problem for the $\epsilon=0$ system \eqref{eq:WhithamNLS} in the focusing ($\kappa=+1$) case can be regarded as a manifestation 
of the well-known modulational instability of the focusing NLS equation.  This instability becomes increasingly severe for $\epsilon\ll 1$ and is asymptotically catastrophic in the semiclassical limit $\epsilon\downarrow 0$.  The modulational instability can lead to unwanted physical effects in systems modeled by the focusing NLS equation.  For example, the Gordon-Haus jitter effect of pulses in optical fiber telecommunications systems is caused at least in part by the modulational instability.  In such situations, physical effects originally neglected to derive a simple model equation (the focusing NLS equation) might play a role in reducing the influence of the instability.  
One effect that can be included perturbatively is nonlinear dispersion, and including such a term
results in the MNLS equation \eqref{eq:MNLS} in place of the focusing NLS equation.  Obviously if $\alpha=0$ the MNLS equation becomes the focusing NLS equation, but for $\alpha>0$ the dynamics can be different, perhaps usefully so.

In terms of the Madelung fluid-dynamical variables the MNLS equation \eqref{eq:MNLS} becomes
\begin{equation}
\frac{\partial \rho_\epsilon}{\partial t} + \frac{\partial}{\partial x}\left(\rho_\epsilon u_\epsilon + \frac{3}{2}\alpha\rho_\epsilon^2\right)=0
\quad\text{and}\quad
\frac{\partial u_\epsilon}{\partial t} + \frac{\partial}{\partial x}\left(\frac{1}{2}u_\epsilon^2-\rho_\epsilon +\alpha\rho_\epsilon u_\epsilon\right) = 
\frac{1}{2}\epsilon^2\frac{\partial F[\rho_\epsilon]}{\partial x},
\label{eq:MNLSMadelung}
\end{equation}
where again $F[\rho]$ is given by \eqref{eq:Fdef}.  The formal limit system (set $\epsilon=0$) is
in this case the \emph{dispersionless MNLS system}
\begin{equation}
\frac{\partial\rho}{\partial t}+\frac{\partial}{\partial x}\left(\rho u+\frac{3}{2}\alpha\rho^2\right)=0\quad\text{and}\quad
\frac{\partial u}{\partial t} +\frac{\partial}{\partial x}\left(\frac{1}{2}u^2-\rho +\alpha\rho u\right)=0.
\label{eq:WhithamMNLS}
\end{equation}
This quasilinear system is obviously an $\alpha$-perturbation of the corresponding elliptic-type dispersionless NLS system
\eqref{eq:WhithamNLS} with $\kappa=+1$.  Unlike the $\alpha=0$ case, the system \eqref{eq:WhithamMNLS} is of
mixed type, being hyperbolic for $Q>0$ and elliptic for $Q<0$,
where
\begin{equation}
Q:=\alpha^2\rho+\alpha u-1.
\end{equation}
This simple calculation suggests that the modulational instability of the focusing NLS equation ($\alpha=0$) can be completely suppressed by nonlinear dispersion ($\alpha>0$) for initial data satisfying the condition $\alpha^2\rho_0(x)+\alpha u_0(x)-1>0$.
Quasilinear systems of mixed type occur frequently in problems of transsonic gas dynamics.  In regions of spacetime where the gas velocity is subsonic (respectively supersonic), the equations are of elliptic (respectively hyperbolic) type.  Throughout this paper we will use this terminology in the context of the mixed-type system \eqref{eq:WhithamMNLS} also.  To make this analogy more obvious, we can define a density-dependent \emph{sound speed} for the dispersionless MNLS system \eqref{eq:WhithamMNLS} by 
\begin{equation}
c(x,t):=\frac{1}{\alpha}-\alpha\rho(x,t),
\end{equation}
and then recalling that $u(x,t)$ has the interpretation of fluid velocity, the condition of hyperbolicity for the system \eqref{eq:WhithamMNLS} becomes the supersonic condition $u(x,t)>c(x,t)$ while that of ellipticity of the system \eqref{eq:WhithamMNLS} becomes the subsonic condition $u(x,t)<c(x,t)$.

The fact that the dispersionless MNLS system \eqref{eq:WhithamMNLS} can be either hyperbolic like the dispersionless defocusing NLS system (\eqref{eq:WhithamNLS} with $\kappa=-1$) or elliptic like the dispersionless focusing NLS system (\eqref{eq:WhithamNLS} with $\kappa=+1$) depending on the sign of $Q$ establishes a qualitative connection between these systems in the semiclassical limit.  However, the connection is far more concrete as was shown in \cite{DiFrancoM08}.  Indeed, let us suppose that $\rho(x,t)$ and $u(x,t)$ are functions satisfying the dispersionless MNLS system \eqref{eq:WhithamMNLS} and the (supersonic) hyperbolicity condition $u(x,t)>c(x,t)$.  Then if new fields $\hat{\rho}(x,t)$ and $\hat{u}(x,t)$ are defined in terms of $\rho(x,t)$ and $u(x,t)$ by the algebraic mapping
\begin{equation}
\hat{\rho}:=\rho Q\quad\text{and}\quad\hat{u}:=u+2\alpha\rho\quad\text{for $Q>0$,}
\label{eq:defocmap}
\end{equation}
it is easy to check that the pair $(\hat{\rho},\hat{u})$ constitutes (after dropping the ``hats'') a solution of the dispersionless defocusing NLS system (\eqref{eq:WhithamNLS} with $\kappa=-1$).  
Similarly, if $\rho(x,t)$ and $u(x,t)$ solve the dispersionless MNLS system \eqref{eq:WhithamMNLS} and satisfy the (subsonic) ellipticity condition $u(x,t)<c(x,t)$,
and if new fields $\hat{\rho}(x,t)$ and $\hat{u}(x,t)$ are instead defined by
\begin{equation}
\hat{\rho}:=-\rho Q\quad\text{and}\quad\hat{u}:=u+2\alpha\rho
\quad\text{for $Q<0$,}
\label{eq:focmap}
\end{equation}
then $(\hat{\rho},\hat{u})$ constitutes a solution of the dispersionless focusing NLS system (\eqref{eq:WhithamNLS} with $\kappa=+1$).  Note that both mappings \eqref{eq:defocmap} and
\eqref{eq:focmap} preserve the positivity of the Madelung fluid density $\rho$.  This simple calculation reveals the remarkable fact that, in the semiclassical limit, the dynamics of the MNLS equation
can be exactly mapped onto either the dynamics of the defocusing NLS equation or those of the focusing NLS equation depending only on the sign of $Q$.

\subsection{Inverse spectral theory.}
\label{sec:inversescattering}
One of our aims in writing this paper is to show how the formal observations made above can be placed on completely rigorous mathematical footing, and also how the interesting transsonic case where $Q$ is sign-indefinite can be analyzed with great precision in the semiclassical limit.  The main tool here is the exact solution of the Cauchy initial-value problem for the MNLS equation \eqref{eq:MNLS} by means of the \emph{inverse spectral transform} specially adapted to that equation.  We also wish to make a comparison with the solution of the NLS equations of both focusing and defocusing types, which is carried out by a better-known inverse spectral transform.
\subsubsection{Inverse spectral transform for MNLS.}
\label{sec:MNLS-IST}
For the reader's convenience, we give here a very brief description (adapted from the appendix of \cite{DiFrancoM08}) of the calculations involved with implementation of the inverse-spectral transform for the MNLS initial-value problem.
Let $\phi_0=\phi(x)$ denote the complex-valued initial condition for the MNLS equation, a function that may depend parametrically on $\epsilon>0$ as in the wavepacket case \eqref{eq:wavepacket} and that is assumed for convenience to be of Schwartz class.  Let $k$ be a complex spectral parameter.
Given $\phi_0$, the \emph{Jost matrices} $\mathbf{J}_\pm(x;k,\epsilon)$ are (uniquely) defined for $k^2\in\mathbb{R}$ to satisfy
the linear differential equation
\begin{equation}
\epsilon\frac{d\mathbf{J}_\pm}{dx}=\mathbf{L}\mathbf{J}_\pm,
\label{eq:scatODE}
\end{equation}
where the coefficient matrix $\mathbf{L}$ is given by
\begin{equation}
\mathbf{L}:=\Lambda\sigma_3 + 2ik\begin{bmatrix}0 & \phi_0(x)\\\phi_0(x)^*&0\end{bmatrix},\quad
\sigma_3:=\begin{bmatrix}
1 & 0 \\ 0 & -1\end{bmatrix},
\end{equation}
and where $\Lambda$ is defined by
\begin{equation}
\Lambda:=-\frac{2i}{\alpha}\left(k^2-\frac{1}{4}\right),
\end{equation}
as well as the asymptotic normalization conditions $\mathbf{J}_\pm(x;k,\epsilon)e^{-\Lambda x\sigma_3/\epsilon}\to\mathbb{I}$ as $x\to\pm\infty$.  The Jost matrix solutions are unimodular for
all $x\in\mathbb{R}$ and $k^2\in\mathbb{R}$, and in particular are fundamental matrix solutions of \eqref{eq:scatODE}.  Since the differential equation  \eqref{eq:scatODE} can have only two linearly independent column vector solutions, the columns of $\mathbf{J}_+(x;k,\epsilon)$ are linear combinations of those of $\mathbf{J}_-(x;k,\epsilon)$ and hence there exists a $2\times 2$ unimodular \emph{scattering matrix} $\mathbf{S}(k,\epsilon)$ defined for $k^2\in\mathbb{R}$ by
\begin{equation}
\mathbf{J}_+(x;k,\epsilon)=\mathbf{J}_-(x;k,\epsilon)\mathbf{S}(k,\epsilon),\quad k^2\in\mathbb{R}.
\label{eq:Sdef}
\end{equation}
In particular, the ratio
\begin{equation}
r(k,\epsilon):=-\frac{S_{12}(k,\epsilon)}{S_{22}(k,\epsilon)},\quad k^2\in\mathbb{R}
\label{eq:rdef}
\end{equation}
is called the \emph{reflection coefficient} corresponding to the potential $\phi_0$; $r(\cdot,\epsilon)$ should be viewed as a kind of transform of $\phi_0(\cdot)$ derived from the behavior of solutions of \eqref{eq:scatODE} as $k$ varies through the continuous spectrum.  By comparing the Jost solutions for $-k$ with those for $k$ it can be shown that $r$ is an odd function of $k$.

There is in general also information about $\phi_0$ encoded in the discrete spectrum of the problem \eqref{eq:scatODE}.
It turns out that
 the second column of $\mathbf{J}_+$ and the first column of $\mathbf{J}_-$ can be analytically continued into the region $\Im\{k^2\}>0$.  
It follows that $S_{22}(k,\epsilon)$ extends analytically to the region $\Im\{k^2\}>0$.
If $k_j\in\mathbb{C}$ with $\Im\{k_j^2\}>0$ is a zero of $S_{22}(k,\epsilon)$, then it is easy to
see that $\mathbf{j}_+^{(2)}(x;k_j,\epsilon)$ and $\mathbf{j}_-^{(1)}(x;k_j,\epsilon)$ are proportional, and hence there exists a nonzero constant $\gamma_j\in\mathbb{C}$ such that
\begin{equation}
\mathbf{j}_+^{(2)}(x;k_j,\epsilon)=\gamma_j\mathbf{j}_-^{(1)}(x;k_j,\epsilon),\quad S_{22}(k_j,\epsilon)=0,\quad \Im\{k_j^2\}>0
\label{eq:proportionality}
\end{equation}
holds for all $x\in\mathbb{R}$.  Such $k=k_j$ are \emph{eigenvalues} or \emph{discrete spectrum} of the differential equation \eqref{eq:scatODE} because it can be shown that the condition  of
proportionality \eqref{eq:proportionality} implies the existence of a solution of \eqref{eq:scatODE} for $k=k_j$ that is exponentially decaying both as $x\to +\infty$ and also as $x\to  -\infty$.

If $\phi_0$ is such that it turns out that $S_{22}(k,\epsilon)$ has a finite (for fixed $\epsilon>0$) number of zeros in its domain of analyticity, all of which are simple zeros (so that $S_{22}'(k_j,\epsilon)\neq 0$, where prime denotes differentiation with respect to $k$), and that $S_{22}(k)$ does not vanish at all for $\Im\{k^2\}=0$, then the corresponding solution $\phi=\phi_\epsilon(x,t)$ of the MNLS equation with
initial condition $\phi_0$ can be obtained from the \emph{scattering data} consisting of the reflection coefficient $r(k,\epsilon)$ for $\Im\{k^2\}=0$ together with the pairs $(k_j,\gamma_j)$ corresponding to the discrete spectrum by the solution of a \emph{matrix Riemann-Hilbert problem}.

The Riemann-Hilbert problem can be formulated to take advantage of a certain symmetry of
the differential equation \eqref{eq:scatODE} under the reflection $k\mapsto -k$.  Thus we introduce
the spectral variable $z=k^2$ and seek a $2\times 2$ matrix $\mathbf{N}(z)=\mathbf{N}(z;x,t,\epsilon)$ satisfying the following conditions ($\mathbf{N}(z):=\mathbf{M}(z^{1/2})$, where $\mathbf{M}(k)$ denotes the matrix solution of the Riemann-Hilbert problem as formulated in the appendix of \cite{DiFrancoM08} and where $z^{1/2}$ denotes the principal branch of the square root).  Let
\begin{equation}
\theta(z;x,t):=-\frac{2}{\alpha}\left(z-\frac{1}{4}\right)x-\frac{4}{\alpha^2}\left(z-\frac{1}{4}\right)^2t,
\end{equation}
and for $z\in\mathbb{R}$ set 
\begin{equation}
s(z,\epsilon):=r(e^{i\pi/4}(-iz)^{1/2},\epsilon)=\begin{cases}r(z^{1/2},\epsilon),\quad& z\ge 0\\ r(i(-z)^{1/2},\epsilon),\quad &z<0.
\end{cases}
\label{eq:sfromr}
\end{equation}
Finally, let $D\subset\mathbb{C}$ denote the finite set $D=\{k_1^2,\dots,k_N^2,k_1^{*2},\dots,k_N^{*2}\}$.
We seek a matrix $\mathbf{N}(z)$ satisfying the following conditions:
\begin{itemize}
\item[]\textit{\textbf{Analyticity:}}  $\mathbf{N}(z)$ is an analytic function of $z$ for $z\in\mathbb{C}\setminus(\mathbb{R}\cup D)$, taking continuous boundary values $\mathbf{N}_\pm(z)$ on
the real line from $\mathbb{C}_\pm$.
\item[]\textit{\textbf{Jump condition:}} The boundary values taken by $\mathbf{N}(z)$ on $\mathbb{R}$
are related as follows:
\begin{equation}
\mathbf{N}_+(z)=\mathbf{N}_-(z)\begin{bmatrix}1 & -s(z,\epsilon)e^{2i\theta(z;x,t)/\epsilon}\\
-s(z,\epsilon)^*e^{-2i\theta(z;x,t)/\epsilon} & 1+|s(z,\epsilon)|^2\end{bmatrix},\quad z\ge 0,
\end{equation}
and
\begin{equation}
\mathbf{N}_+(z)=i^{\sigma_3}\mathbf{N}_-(z)i^{-\sigma_3}\begin{bmatrix}
1 & -s(z,\epsilon)e^{2i\theta(z;x,t)/\epsilon}\\s(z,\epsilon)^*e^{-2i\theta(z;x,t)/\epsilon} & 
1-|s(z,\epsilon)|^2\end{bmatrix},\quad z<0.
\end{equation}
\item[]\textit{\textbf{Point singularities:}}  The matrix $\mathbf{N}(z)$ has simple poles at the points of $D$.
If $z_j=k_j^2\in D$ with $\Im\{z_j\}>0$, then
\begin{equation}
\begin{split}
\mathop{\mathrm{Res}}_{z=z_j}\mathbf{N}(z)&=\lim_{z\to z_j}\mathbf{N}(z)\begin{bmatrix}
0 & -c_je^{2i\theta(z_j;x,t)/\epsilon}\\0 & 0\end{bmatrix}\\
\mathop{\mathrm{Res}}_{z=z_j^*}\mathbf{N}(z)&=\lim_{z\to z_j^*}\mathbf{N}(z)
\begin{bmatrix}0 & 0\\c_j^*e^{-2i\theta(z_j^*;x,t)/\epsilon} & 0\end{bmatrix},
\end{split}
\end{equation}
where $c_j:=-2k_j\gamma_j/S_{22}'(k_j,\epsilon)$.
\item[]\textit{\textbf{Normalization:}}  The matrix $\mathbf{N}(z)$ is normalized to the identity at the origin in the complex plane:  $\mathbf{N}_+(0)=\mathbf{N}_-(0)=\mathbb{I}$.
\end{itemize}
From the solution of this Riemann-Hilbert problem, one extracts the solution $\phi_\epsilon(x,t)$
of the Cauchy initial-value problem for the MNLS equation \eqref{eq:MNLS} with initial data $\phi_\epsilon(x,0)=\phi_0(x)$ from the formula
\begin{equation}
\phi_\epsilon(x,t):=\frac{2}{\alpha}\lim_{z\to\infty}z^{1/2}\frac{N_{12}(z;x,t,\epsilon)}{N_{22}(z;x,t,\epsilon)}.
\label{eq:MNLSreconstruct}
\end{equation}
(Note that $\mathbf{N}(z)=\mathbf{N}_0 + \mathbf{N}_1z^{-1/2} + O(z^{-1})$ as $z\to\infty$,
where $\mathbf{N}_0$ and $\mathbf{N}_1$ are well-defined as functions of $x$ and $t$, and moreover $\mathbf{N}_0$ is diagonal \cite[page 990]{DiFrancoM08} with $\mathrm{det}(\mathbf{N}_0)=1$.)

\subsubsection{Inverse spectral transforms for focusing and defocusing NLS.}
\label{sec:NLS-IST}
As was first shown by Zakharov and Shabat \cite{ZakharovS72}, the initial-value problem for the
NLS equation \eqref{eq:NLS} with Schwartz-class initial data $\phi_0$ can be solved by means of the scattering problem
\begin{equation}
\epsilon\frac{d\mathbf{J}^{\mathrm{ZS},\kappa}_\pm}{dx}=\mathbf{L}^{\mathrm{ZS},\kappa}\mathbf{J}^{\mathrm{ZS},\kappa}_\pm
\label{eq:ZSprob}
\end{equation}
where the coefficient matrix $\mathbf{L}^{\mathrm{ZS},\kappa}$ is given by
\begin{equation}
\mathbf{L}^{\mathrm{ZS},\kappa}:=-i\lambda\sigma_3+\begin{bmatrix}0 & \phi_0(x)\\
-\kappa\phi_0(x)^* & 0\end{bmatrix},
\end{equation}
where $\lambda\in\mathbb{R}$ is a spectral parameter.  Here as before $\mathbf{J}^{\mathrm{ZS},\kappa}_\pm=\mathbf{J}^{\mathrm{ZS},\kappa}_\pm(x;\lambda,\epsilon)$ are Jost matrices satisfying the boundary conditions that
$\mathbf{J}^{\mathrm{ZS},\kappa}_\pm (x;\lambda,\epsilon)e^{i\lambda \sigma_3 x/\epsilon}\to\mathbb{I}$ as $x\to\pm\infty$, and once again one defines a scattering matrix $\mathbf{S}^{\mathrm{ZS},\kappa}(\lambda,\epsilon)$ by the relation
\begin{equation}
\mathbf{J}^{\mathrm{ZS},\kappa}_+(x;\lambda,\epsilon)=\mathbf{J}^{\mathrm{ZS},\kappa}_-(x;\lambda,\epsilon)\mathbf{S}^{\mathrm{ZS},\kappa}(\lambda,\epsilon),\quad \lambda\in\mathbb{R},
\end{equation}
and a reflection coefficient by 
\begin{equation}
r^{\mathrm{ZS},\kappa}(\lambda,\epsilon):=-\frac{S^{\mathrm{ZS},\kappa}_{12}(\lambda,\epsilon)}{S^{\mathrm{ZS},\kappa}_{22}(\lambda,\epsilon)},\quad \lambda\in\mathbb{R}.
\end{equation}
This completes the description of the scattering data associated with the continuous spectrum of the direct scattering problem \eqref{eq:ZSprob}.

If $\kappa=-1$ (the defocusing case), then as an eigenvalue problem with eigenvalue $\lambda$,
the scattering problem \eqref{eq:ZSprob} is self-adjoint for $\phi_0$ Schwartz class, and there can be neither complex eigenvalues nor embedded real ones.  But if $\kappa=+1$ (the focusing case), then selfadjointness is lost and there can be discrete spectrum.  To find the associated scattering
data, note that the second column of $\mathbf{J}_+^{\mathrm{ZS},\kappa}$ and the first column of 
$\mathbf{J}_-^{\mathrm{ZS},\kappa}$ extend analytically into the upper half-plane $\Im\{\lambda\}>0$, and hence so does $S^{\mathrm{ZS},\kappa}_{22}(\lambda,\epsilon)$.  While for $\kappa=-1$ the latter function is bounded away from zero in the closed upper half-plane, if $\kappa=+1$ it may have zeros, which are the discrete eigenvalues of the problem \eqref{eq:ZSprob} for $\kappa=+1$.  If in the latter case $\lambda_j\in\mathbb{C}_+$ is such a zero, then the corresponding columns of the Jost matrices are proportional for $\lambda=\lambda_j$:
\begin{equation}
\mathbf{j}_+^{\mathrm{ZS},\kappa (2)}(x;\lambda_j,\epsilon)=\gamma_j^\mathrm{ZS}
\mathbf{j}_-^{\mathrm{ZS},\kappa (1)}(x;\lambda_j,\epsilon),\quad S_{22}^{\mathrm{ZS},\kappa}(\lambda_j,\epsilon)=0,\quad\Im\{\lambda_j\}>0,\quad \kappa=+1
\end{equation}
for some nonzero constant $\gamma_j^\mathrm{ZS}$.

In the defocusing case, the solution of the Cauchy initial-value problem with Schwartz-class initial data $\phi_0$
can always be obtained from the reflection coefficient alone via the solution of a matrix Riemann-Hilbert problem.  In the focusing case we must additionally assume that $\phi_0$ is such that
the number of zeros of $S^{\mathrm{ZS},\kappa}_{22}(\lambda,\epsilon)$ in the closed upper half-plane is finite, and all are nonreal 
and simple, and we must also include the pairs $(\lambda_j,\gamma_j^{\mathrm{ZS}})$ as part
of the scattering data.  Then, setting
\begin{equation}
\theta^{\mathrm{ZS}}(\lambda;x,t):=-\lambda x - \lambda^2 t,
\end{equation}
and letting $D^{\mathrm{ZS}}:=\{\lambda_1,\dots,\lambda_N,\lambda_1^*,\dots,\lambda_N^*\}$ denote the eigenvalues and their complex conjugates (this is an empty set in the defocusing case),
the Riemann-Hilbert problem is to find a matrix $\mathbf{N}^{\mathrm{ZS},\kappa}(\lambda)=\mathbf{N}^{\mathrm{ZS},\kappa}(\lambda;x,t,\epsilon)$ satisfying the following conditions:
\begin{itemize}
\item[]\textbf{\textit{Analyticity:}} $\mathbf{N}^{\mathrm{ZS},\kappa}(\lambda)$ is an analytic
function of $\lambda$ for $\lambda\in\mathbb{C}\setminus(\mathbb{R}\cup D^\mathrm{ZS})$,
taking continuous boundary values $\mathbf{N}_\pm^{\mathrm{ZS},\kappa}(\lambda)$ from
$\mathbb{C}_\pm$.
\item[]\textbf{\textit{Jump condition:}}  The boundary values taken by $\mathbf{N}^{\mathrm{ZS},\kappa}(\lambda)$ on $\mathbb{R}$ are related as follows:
\begin{equation}
\mathbf{N}_+^{\mathrm{ZS},\kappa}(\lambda)=
\mathbf{N}_-^{\mathrm{ZS},\kappa}(\lambda)\begin{bmatrix}
1 & -r^{\mathrm{ZS},\kappa}(\lambda,\epsilon)e^{2i\theta^\mathrm{ZS}(\lambda;x,t)/\epsilon}\\
-\kappa r^{\mathrm{ZS},\kappa}(\lambda,\epsilon)^*e^{-2i\theta^\mathrm{ZS}(\lambda;x,t)/\epsilon} & 
1+\kappa |r^{\mathrm{ZS},\kappa}(\lambda,\epsilon)|^2\end{bmatrix},\quad z\in\mathbb{R}.
\end{equation}
\item[]\textbf{\textit{Point singularities:}} The matrix $\mathbf{N}^{\mathrm{ZS},\kappa}(\lambda)$
has simple poles at the points of $D^\mathrm{ZS}$ (an empty set for $\kappa=-1$).
If $\lambda_j\in D^\mathrm{ZS}$ with $\Im\{\lambda_j\}>0$, then
\begin{equation}
\begin{split}
\mathop{\mathrm{Res}}_{\lambda=\lambda_j}\mathbf{N}^{\mathrm{ZS},\kappa}(\lambda)&=
\lim_{\lambda\to\lambda_j}\mathbf{N}^{\mathrm{ZS},\kappa}(\lambda)
\begin{bmatrix} 0 & -c_j^\mathrm{ZS}e^{2i\theta^\mathrm{ZS}(\lambda_j;x,t)/\epsilon}\\0 & 0
\end{bmatrix}\\
\mathop{\mathrm{Res}}_{\lambda=\lambda_j^*}\mathbf{N}^{\mathrm{ZS},\kappa}(\lambda)&=
\lim_{\lambda\to\lambda_j^*}\mathbf{N}^{\mathrm{ZS},\kappa}(\lambda)
\begin{bmatrix} 0 & 0\\c_j^{\mathrm{ZS}*}e^{-2i\theta^\mathrm{ZS}(\lambda_j^*;x,t)/\epsilon}& 0
\end{bmatrix},
\end{split}
\end{equation}
where $c_j^\mathrm{ZS}:=-\gamma_j^\mathrm{ZS}/S^{\mathrm{ZS},\kappa\prime}_{22}(\lambda_j,\epsilon)$.
\item[]\textit{\textbf{Normalization:}}  The matrix $\mathbf{N}^{\mathrm{ZS},\kappa}(\lambda)$ is normalized to the identity at infinity in the complex plane:  $\mathbf{N}^{\mathrm{ZS},\kappa}(\lambda)\to\mathbb{I}$ as $\lambda\to\infty$.
\end{itemize}
From the solution of this Riemann-Hilbert problem, one extracts the solution $\phi_\epsilon(x,t)$
of the Cauchy initial-value problem for the NLS equation \eqref{eq:NLS} with initial data $\phi_\epsilon(x,0)=\phi_0(x)$ from the formula
\begin{equation}
\phi_\epsilon(x,t):=2i\lim_{\lambda\to\infty} \lambda N_{12}^{\mathrm{ZS},\kappa}(\lambda;x,t,\epsilon).
\label{eq:NLSreconstruct}
\end{equation}
\subsection{Outline of the paper.}
In this paper we will explain how in the semiclassical limit the MNLS equation can be regarded as a perturbation of either the focusing NLS equation, the defocusing NLS equation, or a mixture of both, depending on the nature of the initial data.  Having already established in \S\ref{sec:formal} a  connection between MNLS and NLS at the level of
 the corresponding dispersionless systems \eqref{eq:WhithamNLS} and \eqref{eq:WhithamMNLS}, we want to indicate how this connection (and especially the crucial role played by the sign of $u(x,t)-c(x,t)$) is manifested at the level of the inverse-spectral transform theory that underpins both the NLS equation \eqref{eq:NLS} and the MNLS equation \eqref{eq:MNLS} as completely integrable partial differential equations.  This is a problem of asymptotic analysis in the semiclassical limit $\epsilon\ll 1$ of the direct (computation of scattering data) and inverse (solution of the matrix Riemann-Hilbert problem) components of the inverse-spectral transform algorithm described in \S\ref{sec:inversescattering}.
 
 In \S\ref{sec:WKB} we present some new asymptotic calculations of the reflection coefficient corresponding to wavepacket initial data for the MNLS equation.  Then, in \S\ref{sec:hyperbolic}
 and \S\ref{sec:elliptic} we interpret these results in the context of, respectively, the conditions of
initial global hyperbolicity and ellipticity of the dispersionless MNLS system \eqref{eq:WhithamMNLS}.  We will show how the type of the latter system at $t=0$ influences
the asymptotic analysis of the Riemann-Hilbert problem of inverse scattering, drawing a comparison with the semiclassical asymptotic analysis of the Riemann-Hilbert problems for the defocusing NLS equation and for the focusing NLS equation respectively.  What will be demonstrated in these sections is quite remarkable: there is a sense in which the direct and inverse spectral transforms for the MNLS initial-value problem with globally hyperbolic (respectively elliptic)
initial data are  \emph{virtually indistinguishable} in the semiclassical limit $\epsilon\ll 1$ from those for the defocusing (respectively focusing) NLS initial-value problem with data obtained from the MNLS initial data via \eqref{eq:defocmap} or \eqref{eq:focmap} as appropriate.  
 
 In a final section we will discuss the interesting case of transsonic initial data for the MNLS equation and the corresponding phenomenon of a ``sonic line'' in the semiclassical MNLS dynamics.  We also indicate how in this situation the asymptotic analysis of the Riemann-Hilbert problem of inverse scattering simultaneously contains features characteristic of both the defocusing and focusing cases.
 
 We are very pleased to be able to contribute our work to this volume of papers in honor of the 85th birthday of Peter Lax.  Peter's interest in dispersionless limits goes back at least to his consideration
 of von Neumann's conjecture that oscillations produced upon shock formation by a finite difference scheme for the integration of a system of partial differential equations of gas dynamics
 should ``on average'' (that is, in the sense of weak convergence as the grid is refined) reproduce the correct entropy shock profile.  In his groundbreaking work in the early 1980's with C. D. Levermore on the zero-dispersion limit of the Korteweg-de Vries equation \cite{LaxL83}, Peter showed that von Neumann's conjecture was false (see \cite{Lax86} for more information about this interesting problem).  

\section{Semiclassical Analysis of the MNLS Reflection Coefficient}
\label{sec:WKB}
For initial data $\phi_0(x)$ of wavepacket form \eqref{eq:wavepacket}, the semiclassical
parameter $\epsilon\ll 1$ enters into the differential equation \eqref{eq:scatODE} in two essential ways:  as a factor multiplying the derivatives with respect to $x$ on the left-hand side, and also in oscillatory exponential factors $e^{\pm iS(x)/\epsilon}$ residing in the off-diagonal elements of the coefficient matrix $\mathbf{L}$ on the right-hand side.  In this section we present some calculations indicating how the limit $\epsilon\to 0$ influences the reflection coefficient $s(z;\epsilon)$ corresponding to wavepacket initial data \eqref{eq:wavepacket} appearing in the Riemann-Hilbert problem of inverse scattering.

Let $\phi_0(x)=\phi_\epsilon(x,0)$ have wavepacket form \eqref{eq:wavepacket}.  We assume that the corresponding functions $\rho_0(x)$ and $u_0(x)$ defined by \eqref{eq:Madelungwavepacket} are such that $\rho_0$ and $u_0'$ are Schwartz-class functions, and that $\rho_0(x)>0$ for all $x\in\mathbb{R}$.  In particular, this condition implies the existence of the limits
\begin{equation}
u_\pm:=\lim_{x\to\pm\infty}u_0(x).
\end{equation}
Letting $S_0:=S(0)$, the relations \eqref{eq:Madelungwavepacket} then imply that
\begin{equation}
S(x)=S_0+\int_0^x u_0(y)\,dy = u_\pm x + S_\pm + o(1),\quad x\to \pm\infty,
\end{equation}
where
\begin{equation}
S_\pm:=S_0 +\int_0^{\pm\infty}\left[u_0(y)-u_\pm\right]\,dy.
\end{equation}
\subsection{Basic setup.}
We will require the Jost matrix solutions for positive real $k$ and positive imaginary $k$ only.  Also, it is convenient to remove the rapidly oscillating exponential factors $e^{\pm i S(x)/\epsilon}$ from the coefficient matrix $\mathbf{L}$ in the differential equation \eqref{eq:scatODE}, so we define related matrices $\mathbf{W}_\pm(x;z,\epsilon)$ for real $z$ by
\begin{equation}
\mathbf{W}_\pm(x;z,\epsilon):= \exp\left(-i\frac{S(x)}{2\epsilon}\sigma_3\right)\cdot
\begin{cases}
\mathbf{J}_\pm(x;\sqrt{z},\epsilon),&\quad z\ge 0\\
\mathbf{J}_\pm(x;i\sqrt{-z},\epsilon),&\quad z <0,
\end{cases}
\label{eq:Wpmdef}
\end{equation}
Thus, the real $z$-axis corresponds to positive real $k$ (for $z>0$) and positive imaginary $k$ (for $z<0$).  The exponential prefactor ensures that the matrices $\mathbf{W}_\pm(x;z,\epsilon)$ satisfy a modification of the differential equation \eqref{eq:scatODE}:
\begin{equation}
2\alpha\epsilon\frac{d\mathbf{W}_\pm}{dx}=i\mathbf{M}\mathbf{W}_\pm,
\label{eq:JostWDE}
\end{equation}
in which the coefficient matrix $\mathbf{M}=\mathbf{M}(x;z)$ is independent of $\epsilon$ and is defined by
\begin{equation}
\mathbf{M}:=\begin{cases}
\begin{bmatrix}-4z+1-\alpha u_0(x) & 4\alpha\sqrt{z}\sqrt{\rho_0(x)}\\
4\alpha\sqrt{z}\sqrt{\rho_0(x)} & 4z-1+\alpha u_0(x)\end{bmatrix},&\quad z\ge 0\\\\
\begin{bmatrix}-4z+1-\alpha u_0(x) & 4i\alpha\sqrt{-z}\sqrt{\rho_0(x)}\\
4i\alpha\sqrt{-z}\sqrt{\rho_0(x)} & 4z-1+\alpha u_0(x)\end{bmatrix},&\quad z<0.
\end{cases}
\label{eq:Mdef}
\end{equation}
The matrices $\mathbf{W}_\pm(x;z,\epsilon)$ are then uniquely determined by the normalization conditions
\begin{equation}
\mathbf{W}_\pm(x;z,\epsilon)\exp\left(i\frac{(4z-1+\alpha u_\pm)x+\alpha S_\pm}{2\alpha\epsilon}\sigma_3\right)
\to
\mathbb{I},\quad x\to\pm\infty,\quad z\in\mathbb{R}.
\label{eq:Wnorm}
\end{equation}

\subsection{WKB formalism.}
The Wenzel-Kramers-Brillouin (WKB) method of approximation can be applied to the differential
equation \eqref{eq:JostWDE} in the asymptotic limit $\epsilon\downarrow 0$.  The form of the WKB
ansatz for a column-vector solution $\mathbf{w}=\mathbf{w}(x;z,\epsilon)$ of \eqref{eq:JostWDE} is 
\begin{equation}
\mathbf{w}=\mathbf{v}(x;z,\epsilon)e^{f(x;z)/(2\alpha\epsilon)},\quad \mathbf{v}(x;z,\epsilon)\sim
\mathbf{v}_0(x;z)+\epsilon\mathbf{v}_1(x;z)+\epsilon^2\mathbf{v}_2(x;z) + \cdots
\label{eq:WKBansatz}
\end{equation}
with the scalar exponent $f$ being independent of $\epsilon$, and the vector $\mathbf{v}(x;z,\epsilon)$
being expanded in an asymptotic series in the parameter $\epsilon$.  Inserting the ansatz into
the differential equation \eqref{eq:JostWDE} and separating powers of $\epsilon$ results in a hierarchy of equations governing the exponent $f$ and the vector coefficients $\mathbf{v}_k(x;z)$:
at leading order we find
\begin{equation}
\mathbf{M}\mathbf{v}_0=-i \frac{df}{dx}\mathbf{v}_0
\label{eq:WKBleading}
\end{equation}
and then at subsequent orders
\begin{equation}
\mathbf{M}\mathbf{v}_k + i \frac{df}{dx}\mathbf{v}_k=-2i\alpha\frac{d\mathbf{v}_{k-1}}{dx},\quad k>0.
\label{eq:WKBhigher}
\end{equation}
Clearly, solving \eqref{eq:WKBleading} with $\mathbf{v}_0$ nowhere vanishing requires that  for each real $x$ and $z$, the characteristic equation 
\begin{equation}
-\left(\frac{df}{dx}\right)^2=(4z-1+\alpha u_0(x))^2+16\alpha^2z\rho_0(x)
\label{eq:characteristic}
\end{equation}
hold true.  We must now distinguish two cases:
\begin{itemize}
\item The oscillatory case.  This corresponds to open intervals of $x$ (depending on $z\in\mathbb{R}$) in which the right-hand side of \eqref{eq:characteristic} is strictly positive.  In this case, $df/dx$ will be taken to be one of two purely imaginary smooth functions of $x$:  we write $df/dx = \pm i\omega(x;z)$ where $\omega(x;z)\in\mathbb{R}$.
\item The exponential case.  This corresponds to complementary open intervals of $x$ in which the right-hand side of \eqref{eq:characteristic} is strictly negative.  In this case, $df/dx$ will be taken to be one
of two purely real smooth functions of $x$:  we write $df/dx = \pm \gamma(x;z)$ where $\gamma(x;z)\in\mathbb{R}$.
\end{itemize}
Since $\rho_0(x)$ vanishes for large $|x|$, if $4z-1+\alpha u_-\neq 0$ then the oscillatory case holds for
$x$ sufficiently negative, while if $4z-1+\alpha u_+\neq 0$ then the oscillatory case holds for $x$ sufficiently positive.
The isolated values of $x$ (depending on $z\in\mathbb{R}$) for which neither the oscillatory nor the exponential case applies because the right-hand side of \eqref{eq:characteristic} vanishes are called
\emph{turning points}.    The WKB ansatz \eqref{eq:WKBansatz} fails near each turning point.

In each oscillatory or exponential interval the derivative $df/dx$ is therefore determined as a smooth 
function of $x$, and then \eqref{eq:WKBleading} further requires that $\mathbf{v}_0$ reside in the nullspace of the singular matrix $\mathbf{M}+i df/dx$.  Since the eigenvalues of $\mathbf{M}$ are distinct, the matrix $\mathbf{M}+i df/dx$ has rank one; let $\mathbf{y}(x;z)$ denote a nonzero vector
in the nullspace chosen arbitrarily, but depending smoothly on $x$ and $z$.  We then see
that $\mathbf{v}_0 = v_0\mathbf{y}$, where $v_0=v_0(x;z)$ is a scalar factor with smooth dependence on $x$ and $z$ to be determined.  Inserting this expression on the right-hand side of \eqref{eq:WKBhigher} for $k=1$ shows that $dv_0/dx$ will need to be determined as a solvability condition for $\mathbf{v}_1$.  With
$dv_0/dx$ so chosen, one then obtains $\mathbf{v}_1$ up to the addition of an arbitrary nullvector of
$\mathbf{M}+i df/dx$, which may be written in the form $v_1\mathbf{y}$ involving a new scalar factor $v_1$, and $dv_1/dx$ will then need to be chosen as a solvability condition for $\mathbf{v}_2$.  The procedure continues in this systematic way for all $k>0$.  Clearly, at each order there is produced
one new integration constant (independent of $x$ and $\epsilon$, but $z$-dependence is allowed).  But these constants may all be combined together, amounting to adding to $\mathbf{v}(x;z,\epsilon)$ a
vector of the form $c(z,\epsilon)\mathbf{y}(x;z)$ where the scalar $c$ has an asymptotic expansion in powers of $\epsilon$.

\subsection{The reflection coefficient in the case of two real turning points.  Barrier tunneling problem.  Generalization for more than two turning points.}
\label{sec:barriertunneling}
Let $\mathbf{w}^{(1)}_\pm$ and $\mathbf{w}^{(2)}_\pm$ denote the first and second columns, respectively, of the normalized matrix solutions $\mathbf{W}_\pm$ of \eqref{eq:JostWDE}.  Since 
$\mathbf{w}_-^{(1)}$ and $\mathbf{w}_-^{(2)}$ are linearly independent solutions of \eqref{eq:JostWDE}
it follows that there are scattering coefficients $T_{12}(z,\epsilon)$ and $T_{22}(z,\epsilon)$ such
that the identity
\begin{equation}
\mathbf{w}_+^{(2)}(x;z,\epsilon)=T_{12}(z,\epsilon)\mathbf{w}_-^{(1)}(x;z,\epsilon)+T_{22}(z,\epsilon)
\mathbf{w}_-^{(2)}(x;z,\epsilon)
\label{eq:scatteringrelation}
\end{equation}
holds.  Note that $T_{jk}(z,\epsilon):=S_{jk}(e^{i\pi/4}(-iz)^{1/2},\epsilon)$ for $z\in\mathbb{R}$,
where the scattering matrix $\mathbf{S}(k,\epsilon)$ is defined by \eqref{eq:Sdef}.  To compute $T_{12}(z,\epsilon)$ and $T_{22}(z,\epsilon)$ we now assume that $z\in\mathbb{R}$ is such that 
\begin{itemize}
\item $4z-1+\alpha u_-\neq 0$,
\item $4z-1+\alpha u_+\neq 0$, and
\item There exist exactly two turning points $x_-(z)<x_+(z)$, and the WKB approximation is in the exponential case for $x_-(z)<x<x_+(z)$.  Note that this condition implies in particular that $z<0$.
\end{itemize}
This situation is analogous to that occurring in one-dimensional quantum scattering problems,
where a particle represented by a wave propagating to the right from $x=-\infty$ encounters a
potential barrier of sufficient height that it lacks the classical energy to penetrate.  In the semiclassical limit $\epsilon\ll 1$, the wave is nearly totally reflected, but an exponentially small
fraction of the incoming wave is transmitted through quantum \emph{tunneling}, and appears at $x=+\infty$ as a small rightward propagating wave.  The mathematical problem that arises there is also one of calculating the both the phase of the reflected wave and also the exponentially small magnitude of the transmitted wave, and so we refer to the calculation we are about to embark upon in the context of the MNLS scattering problem \eqref{eq:scatODE}
a \emph{barrier tunneling problem}.

Our argument is  the following:  $\mathbf{w}_+^{(2)}(x;z,\epsilon)$ is approximated by the WKB
method (oscillatory case) for $x>x_+(z)$.  If we consider $x$ just to the right of $x_+(z)$, we see that
$\mathbf{w}_+^{(2)}(x;z,\epsilon)$ remains bounded as $\epsilon\downarrow 0$, although the phase
varies rapidly.  Now, the WKB method fails in a neighborhood of the turning point $x=x_+(z)$, and to pass through the turning point requires local analysis.   However we will be able to proceed with the formal asymptotics at leading order simply by assuming that the connection problem is solved by constants that are independent of $\epsilon$ (as is the case for simple turning points of the equation $-\epsilon^2 y''(x)+V(x)y(x)=0$, for example).  Under this assumption, the solution $\mathbf{w}^{(2)}_+(x;z,\epsilon)$ is
well-approximated for small $\epsilon>0$ at $x$ just to the left of $x_+(z)$ by a linear combination, with coefficients having magnitudes independent of $\epsilon$, of the two WKB formulae valid in the exponential region $x_-(z)<x<x_+(z)$.  Continuing this approximation to the left, away from the turning point $x_+(z)$, we observe that one of the two WKB formulae becomes exponentially large compared with the other; therefore if $x$ is any point in the interior of the exponential region, $\mathbf{w}^{(2)}_+(x;z,\epsilon)$ is given by 
\begin{equation}
\begin{split}
\mathbf{w}_+^{(2)}(x;z,\epsilon)&=\left(\mathbf{y}_\mathrm{exp}(x;z)+O(\epsilon)\right)\exp\left(\frac{1}{2\alpha\epsilon}\int_x^{x_+(z)}\gamma(y;z)\,dy\right)\\
&\quad\quad{}
+ \text{exponentially small},\quad\epsilon\downarrow 0,\quad x_-(z)<x<x_+(z),
\end{split}
\label{eq:w2exp}
\end{equation}
where
\begin{equation}
\gamma(x;z):=\sqrt{-16\alpha^2z\rho_0(x)-(4z-1+\alpha u_0(x))^2}>0,\quad x_-(z)<x<x_+(z)
\label{eq:gammadef}
\end{equation}
and where $\mathbf{y}_\mathrm{exp}(x;z)$ is a smooth, $\epsilon$-independent eigenvector of $\mathbf{M}$ with eigenvalue $i\gamma(x;z)$.  

To approximate $T_{12}(z,\epsilon)$ and $T_{22}(z,\epsilon)$ for small $\epsilon>0$, we will now construct WKB approximations for the columns of $\mathbf{W}_-(x;z,\epsilon)$ valid in the oscillatory region $x<x_-(z)$.  Let 
\begin{equation}
\sigma_\pm:=\mathrm{sgn}(4z-1+\alpha u_\pm),
\label{eq:sigmapmdef}
\end{equation}
and define a real frequency of
oscillation by
\begin{equation}
\omega(x;z):=\sigma_-\sqrt{16\alpha^2z\rho_0(x)+(4z-1+\alpha u_0(x))^2},\quad x<x_-(z).
\label{eq:omegadef}
\end{equation}
For each appropriate value of $z\in\mathbb{R}$, this quantity has the fixed sign $\sigma_-$ throughout its domain of definition, and $\omega(x;z)\to \omega_-(z)$ as $x\to -\infty$,
where
\begin{equation}
\omega_-(z):=4z-1+\alpha u_-.
\label{eq:omegaminusdef}
\end{equation}
Finally, let $\mathbf{y}_\mathrm{osc}^\pm(x;z)$ denote smooth eigenvectors of $\mathbf{M}(x;z)$ with eigenvalues $\pm\omega(x;z)$ satisfying
\begin{equation}
\lim_{x\to -\infty}\mathbf{y}_\mathrm{osc}^+(x;z)=\begin{bmatrix}0\\1\end{bmatrix}\quad\text{and}\quad
\lim_{x\to -\infty}\mathbf{y}_\mathrm{osc}^-(x;z)=\begin{bmatrix}1\\0\end{bmatrix}.
\end{equation}
Then, the leading-order WKB approximations of the columns of $\mathbf{W}_-(x;z,\epsilon)$ valid 
for $x<x_-(z)$ are
\begin{equation}
\begin{split}
\mathbf{w}^{(1)}_-(x;z,\epsilon)&=\left(\mathbf{y}_\mathrm{osc}^-(x;z)+O(\epsilon)\right)
\exp\left(\frac{i}{\epsilon}\left(C^{(1)}(z)+\frac{1}{2\alpha}\int_x^{x_-(z)}\omega(y;z)\,dy \right)\right)\\
\mathbf{w}^{(2)}_-(x;z,\epsilon)&=\left(\mathbf{y}_\mathrm{osc}^+(x;z)+O(\epsilon)\right)
\exp\left(\frac{i}{\epsilon}\left(C^{(2)}(z)-\frac{1}{2\alpha}\int_x^{x_-(z)}\omega(y;z)\,dy\right)\right)\,,
\end{split}
\label{eq:WKBoscleft}
\end{equation}
where $C^{(1)}(z)$ and $C^{(2)}(z)$ are real constants (independent of $x$ and $\epsilon$) to be determined so
that these formulae are consistent to order $O(\epsilon)$ with the normalization condition \eqref{eq:Wnorm}.  We therefore obtain:
\begin{equation}
C^{(1)}(z)=-C^{(2)}(z)=-\frac{1}{2\alpha}\int_{-\infty}^{x_-(z)}\left[\omega(y;z)-\omega_-(z)\right]\,dy -
\frac{1}{2\alpha}\omega_-(z)x_-(z)-\frac{1}{2}S_-.
\end{equation}
Now we again invoke the assumption that the constants involved in solving the connection problem across the turning point $x=x_-(z)$ are independent of $\epsilon$ at leading order.  Thus, the solution $\mathbf{w}_+^{(2)}(x;z,\epsilon)$ given just to the right of $x_-(z)$ by the exponentially large expression
\eqref{eq:w2exp} will also be exponentially large of the same magnitude for $x$ just to the left of $x_-(z)$, and will be asymptotically represented by a linear combination, with constants of the same exponentially large magnitude, of the two oscillatory WKB formulae \eqref{eq:WKBoscleft} valid for $x<x_-(z)$.  This line of
reasoning leads to the formulae
\begin{equation}
\begin{split}
T_{12}(z,\epsilon)&=\exp\left(\frac{1}{2\alpha\epsilon}\int_{x_-(z)}^{x_+(z)}\gamma(y;z)\,dy\right)
e^{-iC^{(1)}(z)/\epsilon}
\cdot O(1),\quad \epsilon\downarrow 0\\
T_{22}(z,\epsilon)&=\exp\left(\frac{1}{2\alpha\epsilon}\int_{x_-(z)}^{x_+(z)}\gamma(y;z)\,dy\right)
e^{-iC^{(2)}(z)/\epsilon}\cdot O(1),\quad\epsilon\downarrow 0.
\end{split}
\end{equation}
The reflection coefficient $s(z,\epsilon)$ relevant for the inverse-scattering problem is defined in terms of $S_{12}(k,\epsilon)$ and $S_{22}(k,\epsilon)$ by \eqref{eq:rdef} and \eqref{eq:sfromr}.
From the above asymptotic formulae for $T_{12}$ and $T_{22}$ it then follows that
\begin{equation}
\begin{split}
\Phi(z):=\lim_{\epsilon\downarrow 0}\left(-i\epsilon\log(s(z,\epsilon))\right) &= C^{(2)}(z)-C^{(1)}(z) \\ &=
\frac{1}{\alpha}\int_{-\infty}^{x_-(z)}\left[\omega(y;z)-\omega_-(z)\right]\,dy +\frac{1}{\alpha}\omega_-(z)x_-(z) +S_-.
\end{split}
\label{eq:PhidefWKB}
\end{equation}
An exact identity within the scattering theory that is a consequence of antiholomorphic spectral symmetry and unimodularity of the scattering matrix is 
\begin{equation}
|T_{22}(z,\epsilon)|^2-|T_{12}(z,\epsilon)|^2=1,\quad z<0,
\end{equation}
and this together with the asymptotic formulae for $T_{12}$ and $T_{22}$ implies that
\begin{equation}
\tau(z):=-\lim_{\epsilon\downarrow 0}\left(\epsilon\log(1-|s(z,\epsilon)|^2)\right) = \frac{1}{\alpha}
\int_{x_-(z)}^{x_+(z)}\gamma(y;z)\,dy
\label{eq:tauWKB}
\end{equation}
whenever $z$ is a (necessarily negative) value of the spectral parameter for which there
are exactly two real turning points.  To summarize, for such $z<0$,
the reflection coefficient $s(z,\epsilon)$ has modulus exponentially close to $1$ (as measured by $\tau(z)>0$) and real fast phase approximated by $\epsilon^{-1}\Phi(z)$.

The functions $\Phi(z)$ and $\tau(z)$ can be related by analytic continuation in the case that the
coefficient functions $u_0$ and $\rho_0$ are analytic.  Indeed, suppose this is the case, and that $z_0$ is a real value of $z$ at which the real turning points $x_\pm(z)$ coalesce from below ($z<z_0$),
so that in particular $x_+(z_0)=x_-(z_0)$.  For $z<z_0$, $x_\pm(z)$ are two branches of the inverse of an analytic function $z=z(x)$ satisfying $z'(x_\pm(z_0))=0$.  If we presume the generic condition $z''(x_\pm(z_0))<0$, then it follows that $x_\pm(z)$ behave locally like square roots
of $z_0-z$.  In particular, a full circuit of $z$ about $z_0$ in the positive (negative) sense results
in a half-circuit of $x_\pm(z)$ about $x_\pm(z_0)$ in the same sense, and this also leads to the
permutation of the two turning points:  $x_\pm(z_0+(z-z_0)e^{2\pi i}) = x_\mp(z)$ holds for
$z<z_0$, as does $x_\pm(z_0+(z-z_0)e^{-2\pi i})=x_\mp(z)$.  For $z<z_0$, we choose $x_0<x_-(z)$ and write $\Phi(z)$ in the form
\begin{equation}
\Phi(z)=\frac{1}{\alpha}\int_{-\infty}^{x_0}[\omega(y;z)-\omega_-(z)]\,dy +\frac{1}{\alpha}x_0\omega_-(z) +S_-+\frac{1}{\alpha}\int_{x_0}^{x_-(z)}\omega(y;z)\,dy,\quad z<z_0.
\end{equation}
The first two terms on the right-hand side are analytic in $z$ at $z=z_0$, and hence will return
to the same value upon a complete circuit of $z$ about $z_0$ in either the positive or negative
sense.  The third term, however, experiences monodromy upon such an analytic continuation, 
because the turning points are exchanged:
\begin{equation}
\frac{1}{\alpha}\int_{x_0}^{x_-(z_0+(z-z_0)e^{\pm 2\pi i})}\omega(y;z_0+(z-z_0)e^{\pm 2\pi i})\,dy = 
\frac{1}{\alpha}\int_{x_0}^{x_-(z)}\omega(y;z)\,dy +\frac{1}{\alpha}\int_{x_-(z)}^{x_+(z)}
\omega_\mp(y;z)\,dy,
\end{equation}
where for $x_-(z)<y<x_+(z)$,
\begin{equation}
\omega_\pm(y;z):=\lim_{\delta\downarrow 0}\omega(y\pm i\delta;z)=\mp i\sigma_-\gamma(y;z).
\end{equation}
It therefore follows that
\begin{equation}
\Phi(z_0+(z-z_0)e^{\pm 2\pi i})=\Phi(z)\pm i\sigma_-\tau(z),\quad z<z_0.
\label{eq:Phimonodromy}
\end{equation}
It is easy to check that exactly the same formula holds true for analytic continuation around
a point $z_0\in\mathbb{R}$ at which $x_\pm(z)$ coalesce like square roots from above ($z>z_0$).
Even simpler reasoning (write $\tau$ as a loop integral around a complete branch cut between $x_-(z)$ and $x_+(z)$) produces the continuation formula
\begin{equation}
\tau(z_0+(z-z_0)e^{\pm 2\pi i})=\tau(z)
\label{eq:taumonodromy}
\end{equation}
for analytic continuation about both types of branching points $z_0$.  

Finally, we note that the arguments presented in this section also carry over to the case when
there exist  arbitrarily many simple turning points  (necessarily an even number).  Indeed, if $x_-(z)$ is simply reinterpreted as the left-most turning point, then the formula \eqref{eq:PhidefWKB} for the
phase $\Phi(z)$ of the reflection coefficient still holds, and the formula \eqref{eq:tauWKB} for $\tau(z)$ needs only to be modified by replacing the integration interval by the union of intervals
in which the exponential case holds, that is, where the radicand of $\gamma(x;z)$ as defined by
\eqref{eq:gammadef} is positive.

\subsection{The reflection coefficient  in the absence of real turning points.  Above-barrier reflection.}
Now suppose that $z\in\mathbb{R}$ is a value for which the right-hand side of 
\eqref{eq:characteristic} is uniformly bounded away from zero for $x\in\mathbb{R}$.  In fact the right-hand side
of \eqref{eq:characteristic} is necessarily positive in this case, since $\rho_0(x)\to 0$ as $|x|\to\infty$.  Therefore we are in the oscillatory case for the WKB method over the whole real $x$-axis, and in particular, for smooth $\rho_0$ and $u_0$ the function $\omega(x;z)$ is well-defined by \eqref{eq:omegadef} as a smooth function of  $x\in\mathbb{R}$ whenever $z\in\mathbb{R}$.  Moreover,
it can be shown that the WKB ansatz is uniformly valid for $x\in\mathbb{R}$ to all orders of accuracy.
In other words, for each choice of sign in the formula $df/dx=\pm i\omega(x;z)$ there exists a one-dimensional subspace of true solutions $\mathbf{w}(x;z,\epsilon)$ for which the asymptotic expansion
\eqref{eq:WKBansatz} holds uniformly for $x\in\mathbb{R}$; the magnitude of the difference between $\mathbf{v}(x;z,\epsilon):=\mathbf{w}(x;z,\epsilon)e^{-f(x;z,\epsilon)/(2\alpha\epsilon)}$ and the partial sum
$\mathbf{v}_0(x;z)+\epsilon\mathbf{v}_1(x;z)+\cdots+\epsilon^N\mathbf{v}_N(x;z)$ has a maximum
value over $x\in\mathbb{R}$ that is $O(\epsilon^{N+1})$ as $\epsilon\downarrow 0$.  

The absence of turning points in a quantum scattering problem occurs when a classical particle incident on a potential barrier from $x=-\infty$ has sufficient energy to penetrate the barrier (or propagate ``above'' the barrier) and arrive at $x=+\infty$.  Quantum mechanically, however, when the particle is represented as a right-going incident wave, the barrier still causes the generation of an exponentially small reflected wave (and similarly, the transmitted wave is attenuated by an exponentially small fraction).  In the context of the scattering problem \eqref{eq:scatODE} for the MNLS equation, the ``reflection coefficient'' $s(z,\epsilon)$ can correspond either to the magnitude of the physical reflection coefficient or to its reciprocal, depending on a monodromy index associated with the coefficients representing the barrier.  Despite this difference, we refer to the calculation we are about to embark upon as a problem of \emph{above-barrier reflection}.

Let $N_\pm=N_\pm(x;z)$ be given by
\begin{equation}
N_\pm:=\begin{cases}
\sqrt{16\alpha^2z\rho_0(x) +[4z-1+\alpha u_0(x)\pm\omega(x;z)]^2},&\quad z>0\\
\sqrt{\pm\left(16\alpha^2z\rho_0(x)+[4z-1+\alpha u_0(x)\pm\omega(x;z)]^2\right)},&\quad z<0.
\end{cases}
\end{equation}
In both cases the square roots are taken to be positive.  
(It is obvious that the radicand appearing in the definition of $N_\pm$ is positive if $z> 0$.   On the other hand, if $z<0$, we have
\begin{equation}
16\alpha^2z\rho_0(x)+[4z-1+\alpha u_0(x)\pm\omega(x;z)]^2 = 2\omega(x;z)\left[\omega(x;z) \pm (4z-1+\alpha u_0(x))\right],
\label{eq:rawradicand}
\end{equation}
and since $z<0$, $0<|\omega(x,z)|<|4z-1+\alpha u_0(x)|$ so the sign of \eqref{eq:rawradicand} is the same as that of $\omega(x;z)$, namely $\sigma_-$, times that of $\pm (4z-1+\alpha u_0(x))$.  But since we are in the oscillatory case for all $x\in\mathbb{R}$ by assumption, $z<0$ implies that $4z-1+\alpha u_0(x)$ is bounded away from zero for $x\in\mathbb{R}$, and therefore the latter sign can be taken from the limiting value at $x=-\infty$, namely $\mathrm{sgn}(\pm(4z-1+\alpha u_0(x)))=\pm\sigma_-$, so the radicand is again positive.)
An eigenvector of $\mathbf{M}$ with eigenvalue $\pm\omega(x;z)$ assumed real is given by
\begin{equation}
\mathbf{y}_\pm(x;z) = \frac{1}{N_\pm}\begin{cases}\begin{bmatrix} 4\alpha\sqrt{z}\sqrt{\rho_0(x)}\\
4z-1+\alpha u_0(x)\pm \omega(x;z)\end{bmatrix},&\quad z> 0\\\\
\begin{bmatrix}4i\alpha\sqrt{-z}\sqrt{\rho_0(x)}\\4z-1+\alpha u_0(x)\pm\omega(x;z)\end{bmatrix},&\quad z<0.
\end{cases}
\end{equation}
It is then a straightforward calculation to confirm that by choice of the normalization constants $N_\pm$, we
have
\begin{equation}
\mathbf{y}_\pm(x;z)^\mathsf{T}\mathbf{y}_\mp(x;z)=0\quad\text{and}\quad
\mathbf{y}_\pm(x;z)^\mathsf{T}\mathbf{y}_\pm(x;z)=\begin{cases}1, &\quad z>0\\
\pm 1, &\quad z<0.
\end{cases}
\label{eq:ynorm}
\end{equation}
We also record here the asymptotic values of $\mathbf{y}_\pm(x;z)$ as $|x|\to\infty$:  if $z>0$,
we have
\begin{equation}
\lim_{x\to -\infty}\mathbf{y}_+(x;z)=\begin{bmatrix}0\\1\end{bmatrix}\quad\text{and}\quad
\lim_{x\to -\infty}\mathbf{y}_-(x;z)=\begin{bmatrix}1\\0\end{bmatrix},
\end{equation}
while
\begin{equation}
\lim_{x\to +\infty}\mathbf{y}_+(x;z)=\begin{cases}
\displaystyle\begin{bmatrix}0\\1\end{bmatrix},&\quad\sigma_+=\sigma_-\\\\
\displaystyle\begin{bmatrix}1\\0\end{bmatrix},&\quad\sigma_+\neq\sigma_-
\end{cases}\quad\text{and}\quad
\lim_{x\to +\infty}\mathbf{y}_-(x;z)=\begin{cases}
\displaystyle\begin{bmatrix}1\\0\end{bmatrix},&\quad\sigma_+=\sigma_-\\\\
\displaystyle\begin{bmatrix}0\\1\end{bmatrix},&\quad\sigma_+\neq\sigma_-.
\end{cases}
\end{equation}
On the other hand, if $z<0$, 
then we necessarily have $\sigma_+=\sigma_-$ under the assumption that there are no real turning points, and so
\begin{equation}
\lim_{|x|\to\infty}\mathbf{y}_+(x;z)=\begin{bmatrix}0\\1\end{bmatrix}\quad\text{and}\quad
\lim_{|x|\to\infty}\mathbf{y}_-(x;z)=\begin{bmatrix}i\\0\end{bmatrix}.
\end{equation}

Now let $\mathbf{Y}(x;z):=(\mathbf{y}_+(x;z),\mathbf{y}_-(x;z))$ be the eigenvector matrix.  From
\eqref{eq:ynorm} it follows that
\begin{equation}
\mathbf{Y}(x;z)^{-1}=\begin{cases}\mathbf{Y}(x;z)^\mathsf{T},&\quad z>0\\
\sigma_3\mathbf{Y}(x;z)^\mathsf{T},&\quad z<0.
\end{cases}
\label{eq:Yinverse}
\end{equation}
From the identity $\mathbf{Y}(x;z)^{-1}\mathbf{Y}(x;z)=\mathbb{I}$ we get by differentiation
\begin{equation}
\mathbf{0}=\frac{d\mathbf{Y}(x;z)^{-1}}{dx}\mathbf{Y}(x;z)+\mathbf{Y}(x;z)^{-1}\frac{d\mathbf{Y}(x,z)}{dx},
\end{equation}
so combining with \eqref{eq:Yinverse} we see that regardless of whether $z>0$ or $z<0$,
\begin{equation}
\mathbf{Y}(x;z)^{-1}\frac{d\mathbf{Y}(x;z)}{dx} = \begin{bmatrix} 0 & h(x;z)\\-h(x;z) & 0\end{bmatrix}
\end{equation}
for some smooth function $h(x;z)$.  We would like to use this information to calculate the
Jost solution $\mathbf{w}^{(2)}_+(x;z,\epsilon)$ in the limit $\epsilon\downarrow 0$.  According
to \eqref{eq:Wnorm}, we should choose 
\begin{equation}
f(x;z)=i\alpha S_+ +i\sigma_+\sigma_-\omega_+(z)x-i\sigma_+\sigma_-\int_x^{+\infty}\left[\omega(y;z)-\omega_+(z)\right]\,dy\quad\text{to obtain $\mathbf{w}^{(2)}_+(x;z,\epsilon)$}.
\label{eq:dfdxvalue}
\end{equation}
Here,
\begin{equation}
\omega_+(z):=\lim_{x\to +\infty}\omega(x;z)=\sigma_+\sigma_-(4z-1+\alpha u_+).
\end{equation}

It is now easy to carry out the calculation of the WKB approximation to all orders, by asymptotically diagonalizing the
system governing $\mathbf{v}(x;z,\epsilon):=\mathbf{w}(x;z,\epsilon)e^{-f(x;z)/(2\alpha\epsilon)}$ with the use of the substitution $\mathbf{v}(x;z,\epsilon)=\mathbf{Y}(x;z)\mathbf{z}(x;z,\epsilon)$.  In the case $\sigma_+=\sigma_-$ so that $df/dx=i\omega(x;z)$, 
 the system satisfied exactly by $\mathbf{z}=(\zeta_1,\zeta_2)^\mathsf{T}$ consists of the equations
\begin{equation}
\begin{split}
0&=-2i\alpha\epsilon\left(h(x;z)\zeta_{2}(x;z,\epsilon) +\frac{d\zeta_{1}}{dx}(x;z,\epsilon)\right)\\
-2\omega(x;z)\zeta_{2}(x;z,\epsilon) &= -2i\alpha\epsilon\left(-h(x;z)\zeta_{1}(x;z,\epsilon) +\frac{d\zeta_{2}}{dx}(x;z,\epsilon)\right).
\end{split}
\label{eq:zetasystemplus}
\end{equation}
On the other hand, in the case $\sigma_+\neq\sigma_-$ (implying $z>0$) so that $df/dx=-i\omega(x;z)$, the equations become
\begin{equation}
\begin{split}
2\omega(x;z)\zeta_1(x;z,\epsilon)&=-2i\alpha\epsilon\left(h(x;z)\zeta_{2}(x;z,\epsilon) +\frac{d\zeta_{1}}{dx}(x;z,\epsilon)\right)\\
0 &= -2i\alpha\epsilon\left(-h(x;z)\zeta_{1}(x;z,\epsilon) +\frac{d\zeta_{2}}{dx}(x;z,\epsilon)\right).
\end{split}
\label{eq:zetasystemminus}
\end{equation}
Note that since $\rho$ and $u'$ are Schwartz class,  for all $z$ under consideration, $\omega(x;z)^{-1}$ is bounded and infinitely differentiable, and $h(x;z)$ is Schwartz class.  

To solve \eqref{eq:zetasystemplus} we eliminate $\zeta_1$ using the first equation and building in
the correct boundary condition implied by \eqref{eq:Wnorm} at $x=+\infty$:
\begin{equation}
\zeta_1(x;z,\epsilon)=1+\int_{x}^{+\infty}h(y;z)\zeta_2(y;z,\epsilon)\,dy,
\end{equation}
and therefore rewrite the second equation of \eqref{eq:zetasystemplus} in the form
\begin{equation}
\zeta_2(x;z,\epsilon) = \frac{i\alpha\epsilon}{\omega(x;z)}\left[\frac{d\zeta_2}{dx}(x;z,\epsilon)-h(x;z)-h(x;z)\int_{x}^{+\infty}h(y;z)\zeta_2(y;z,\epsilon)\,dy\right].
\end{equation}
Iterating this equation starting with the initial guess $\zeta_2^0(x;z,\epsilon)\equiv 0$ produces the partial sums of the asymptotic power series for $\zeta_2(x;z;\epsilon)$:
\begin{equation}
\zeta_2(x;z,\epsilon)\sim\sum_{n=1}^\infty\zeta_{2,n}(x;z)\epsilon^n,\quad\epsilon\downarrow 0.
\end{equation}
It is straightforward to confirm by induction
that every term in this power series is a Schwartz-class function of $x$. 

In a completely analogous
fashion we solve \eqref{eq:zetasystemminus} 
obtaining the asymptotic power series
\begin{equation}
\zeta_1(x;z,\epsilon)\sim\sum_{n=1}^\infty\zeta_{1,n}(x;z)\epsilon^n,\quad\epsilon\downarrow 0
\end{equation}
in which each term is a Schwartz-class function of $x$.  

The uniform validity of
the WKB approximation in the absence of turning points means that we may interchange limits to 
deduce the behavior of $\mathbf{w}_+^{(2)}(x;z,\epsilon)$ as $x\to -\infty$.  Now, the coefficients
$T_{12}(z,\epsilon)$ and $T_{22}(z,\epsilon)$ in \eqref{eq:scatteringrelation} may be expressed in
terms of Wronskians as
\begin{equation}
T_{12}(z,\epsilon)=\det\left(\mathbf{w}_+^{(2)}(x;z,\epsilon),\mathbf{w}_-^{(2)}(x;z,\epsilon)\right)\quad
\text{and}\quad
T_{22}(z,\epsilon)=\det\left(\mathbf{w}_-^{(1)}(x;z,\epsilon),\mathbf{w}_+^{(2)}(x;z,\epsilon)\right)
\end{equation}
and the right-hand sides are actually independent of $x$, so we may take the limit $x\to-\infty$
and use the asymptotics given by \eqref{eq:Wnorm} to write these formulae exactly in the form
\begin{equation}
\begin{split}
T_{12}(z,\epsilon)&=\lim_{x\to -\infty}e^{i(\omega_-(z)x+\alpha S_-)/(2\alpha\epsilon)}w^{(2)}_{+1}(x;z,\epsilon)\\
T_{22}(z,\epsilon)&=\lim_{x\to -\infty}e^{-i(\omega_-(z)x+\alpha S_-)/(2\alpha\epsilon)}w^{(2)}_{+2}(x;z,\epsilon).
\end{split}
\end{equation}
Substituting the WKB asymptotics for the components of $\mathbf{w}_+^{(2)}(x;z,\epsilon)$ we find differing behavior of $s(z,\epsilon)$ depending on the sign of the \emph{monodromy index} $m:=\sigma_+\sigma_-$:
\begin{itemize}
\item If $m=+1$, (that is, $\sigma_+=\sigma_-$), then $T_{12}(z,\epsilon) = o(\epsilon^n)$ for every integer $n$ while
$T_{22}(z,\epsilon)$ is bounded away from zero in the limit $\epsilon\downarrow 0$.  The reflection 
coefficient $s(z,\epsilon)=-T_{12}(z,\epsilon)/T_{22}(z,\epsilon)$ is therefore small beyond all
orders in $\epsilon$.
\item If $m=-1$ or $\sigma_+\neq\sigma_-$ (which implies in particular $z>0$ in the absence of real turning points), then $T_{22}(z,\epsilon)=o(\epsilon^n)$ for every integer $n$ while $T_{12}(z,\epsilon)$ is bounded away from zero as $\epsilon\downarrow 0$.  In this case, the \emph{reciprocal} of the reflection coefficient is small beyond
all orders in $\epsilon$, and we should expect the reflection coefficient to be exponentially large.
\end{itemize}
In the case that $s(z,\epsilon)$ is small ($\sigma_+=\sigma_-$), the jump matrix in the Riemann-Hilbert problem of inverse scattering is well-approximated by the identity, and no further information about the reflection coefficient is really required.  However when $s(z,\epsilon)$ is large ($\sigma_+\neq \sigma_-$), further asymptotic analysis of the Riemann-Hilbert problem requires a leading-order formula for the reflection coefficient.  This is a challenging problem of exponential asymptotics that appears to require the analyticity of $\rho_0$ and $u_0$ with respect to $x$ as the
most effective methods rely on analytic continuation of the WKB solutions from the real line into the complex $x$-plane.  If $\rho_0$ and $u_0$ are analytic functions of $x$, then given $z\in\mathbb{R}$ for which there are no real turning points we may instead seek \emph{complex} turning points, that is, complex roots $x=x(z)\in\mathbb{C}$ of the right-hand side of \eqref{eq:characteristic}.  These obviously come in complex-conjugate pairs for real $z$.  If the right-hand side of \eqref{eq:characteristic} is analytic in a strip of the complex $x$-plane containing
the pair of complex-conjugate turning points closest to the real $x$-axis, then one can show that
the formula \eqref{eq:PhidefWKB} continues to hold for the phase $\Phi(z)$ of the reflection coefficient, in which $x_-(z)$ is interpreted as the turning point in the upper (respectively, lower) half-plane
for $\sigma_-=-1$ (respectively, $\sigma_-=+1$), and the path of integration is confined to the strip of analyticity.
It is then clear that $\Re\{i\Phi(z)\}>0$, making the reflection coefficient exponentially large.

For $z$ real and of sufficiently large absolute value, it is obvious that there exist no real turning points,
and also that $\sigma_+=\sigma_-$, making the reflection coefficient small beyond all orders.  Also, it is easy to see that $\sigma_+\neq \sigma_-$ exactly for those $z\in\mathbb{R}$ lying in the
interval $I$ given by
\begin{equation}
I:=\left(\frac{1}{4}(1-\alpha\max\{u_+,u_-\}),\frac{1}{4}(1-\alpha\min\{u_+,u_-\})\right).
\label{eq:Idefine}
\end{equation}
Note, however, that $z\in I$ does not guarantee the absence of real turning points; for example for each $z\in I$ with $z\le 0$ there necessarily exists
at least one real turning point (in fact at least two if they are simple).

\subsection{Discrete spectrum.}  Locating eigenvalues for the scattering problem \eqref{eq:scatODE} is also a difficult problem of exponential asymptotics.  The discrete spectrum
(at least the non-embedded part) is nonreal, and it is easy to see from the characteristic equation \eqref{eq:characteristic} that in order to have exponential decay of the eigenfunction as $x\to\pm\infty$ it is necessary to have $\Im\{z\}\neq 0$.  To construct an eigenfunction using the WKB method, one needs to be in the exponential case for sufficiently large $|x|$ and to connect
the two regions of decay through a central oscillatory region via two (or possibly more) turning points.  The problem is that for typical nonreal $z$, there exist no real turning points at all.  One
way to get around this difficulty is to further assume that the initial data $(\rho_0(x),u_0(x))$ are real analytic functions of $x$ admitting analytic continuation into the complex $x$-plane.  In this situation, it is frequently possible to find pairs of \emph{complex} turning points $x_\pm(z)\in\mathbb{C}$ and
to construct an approximate eigenfunction that is oscillatory along a certain curve connecting the turning points and that is exponentially decaying away from the turning points in basins that contain the two points at infinity on the real $x$-axis.  The condition that a curve exists in the complex $x$-plane connecting the turning points along which the real part of the WKB exponent is not changing determines a curve in the complex $z$-plane near which one expects to find $O(\epsilon^{-1})$ eigenvalues approximately determined by a Bohr-Sommerfeld integral condition that quantizes the phase increment between the turning points.  This sort of calculation is described in the context of the nonselfadjoint Zakharov-Shabat eigenvalue problem in \cite{Miller01}.  

If the reflection coefficient $s(z,\epsilon)=-T_{12}(z,\epsilon)/T_{22}(z,\epsilon)$ has a meromorphic continuation into the upper half $z$-plane
from the neighborhood of some $z_0\in\mathbb{R}$ (which is really a statement about the scattering coefficient $T_{12}(z,\epsilon)$, since $T_{22}(z,\epsilon)$ is analytic in the upper half-plane), then it will have simple poles at the eigenvalues $z_j=k_j^2$, and the corresponding proportionality constant $\gamma_j$ defined by \eqref{eq:proportionality} is given by $\gamma_j = T_{12}(z_j,\epsilon)$.  This observation can allow the direct computation of eigenvalues to be avoided in some situations.

\section{Supersonic Initial Data}
\label{sec:hyperbolic}
Here we consider the solution of the MNLS equation \eqref{eq:MNLS} by the inverse-spectral transform in the case that the initial data is globally supersonic for the dispersionless limit system \eqref{eq:WhithamMNLS}.  That is, we assume the initial data to satisfy the condition that
$Q(x):=\alpha^2\rho_0(x)+\alpha u_0(x)-1>0$ for all $x\in\mathbb{R}$.

The discriminant of the quadratic polynomial in $z$ on the right-hand side of the characteristic equation \eqref{eq:characteristic} is $256\alpha^2\rho_0(x)Q(x)$.  Hence
for supersonic initial data there exist two real roots $z=z_\pm(x)$ defined for all $x\in\mathbb{R}$,
with $z_+(x)>z_-(x)$.  Also, directly from the quadratic formula one checks that the supersonic condition and the positivity of $\rho_0(x)$ imply that $z_+(x)\le 0$ for all $x\in\mathbb{R}$, with equality only for those $x$ for which $\alpha u_0(x)=1$.  In general, the locus of complex values of $z$ parametrized by $x\in\mathbb{R}$ for which the right-hand side of  the characteristic equation \eqref{eq:characteristic} vanishes is called the \emph{turning point curve}.  So we see that for globally supersonic initial data the turning point curve is more specifically a real curve.  

Due to the presence of the additional factor of $\rho_0(x)$ in the discriminant, we also have
that $z_+(x)$ and $z_-(x)$ coalesce as $x\to\pm\infty$ to the value $\tfrac{1}{4}(1-\alpha u_\pm)$.  Therefore, upon setting
\begin{equation}
z_\mathrm{L}:=\min_{x\in\mathbb{R}}z_-(x)\quad\text{and}\quad
z_\mathrm{R}:=\max_{x\in\mathbb{R}}z_+(x)\le 0,
\end{equation}
we have the following:
\begin{itemize}
\item For $z_\mathrm{L}<z<z_\mathrm{R}$, there exist real turning points (and by Sard's Theorem the turning points are simple for $z$ in a subset of this interval of full measure).  For such $z$ the reflection coefficient $s(z,\epsilon)$ has modulus exponentially close to $1$ as measured by 
the function $\tau(z)>0$ defined by \eqref{eq:tauWKB}, and has approximate real-valued phase $\epsilon^{-1}\Phi(z)$ where $\Phi(z)$ is defined by \eqref{eq:PhidefWKB}.
\item For $z<z_\mathrm{L}$ and $z>z_\mathrm{R}$, there exist no real turning points at all,
and since $z_\mathrm{L}\le \tfrac{1}{4}(1-\alpha u_\pm)\le z_\mathrm{R}$, we have $I\subset [z_\mathrm{L},z_\mathrm{R}]$ where $I$ is the interval defined by \eqref{eq:Idefine}.  Consequently,
the reflection coefficient $s(z,\epsilon)$ is small beyond all orders for $z\not\in [z_\mathrm{L},z_\mathrm{R}]$.
\end{itemize}
Moreover, one of the general results of \cite{DiFrancoM08} is a hard estimate on the discrete spectrum of the scattering problem \eqref{eq:scatODE} that confines the eigenvalues to the ``shadow'' cast by the turning point curve in the complex $z$-plane by light projected from infinity along vertical lines toward the real axis, up to an error proportional to $\epsilon$.  In the present case, the turning point curve itself is real, and hence it is its own shadow.  As this shadow has no intersection with the open upper half-plane, the conclusion is that, at least at the level of the semiclassical approximation, there are no eigenvalues at all for supersonic initial conditions of the type considered here.

These results should be compared with those for the direct scattering problem for the inverse-spectral transform adapted to the defocusing NLS equation (\eqref{eq:NLS} with $\kappa=-1$).
The analogue of \eqref{eq:scatODE} in this case is the self-adjoint eigenvalue problem \eqref{eq:ZSprob} in the case $\kappa=-1$.  The continuous spectrum for this problem is the real axis of the spectral parameter (in this case, $\lambda$), just as in the case of the scattering problem \eqref{eq:scatODE} for MNLS.  By self-adjointness, there can be no nonreal eigenvalues (discrete spectrum) whatsoever, which is an exact version of the semiclassical asymptotic ``shadow'' estimate for the MNLS scattering problem \eqref{eq:scatODE} with supersonic initial data.  

Our expectation is that the semiclassical defocusing NLS dynamics, \emph{after being written in terms of the MNLS fields $\rho_\epsilon$ and $u_\epsilon$ via the explicit mapping \eqref{eq:defocmap}}, should resemble the semiclassical MNLS dynamics in the case of supersonic initial data.  In other words, we begin with functions $\rho_0(x)$ and $u_0(x)$ satisfying the condition $Q(x)>0$ for all $x\in\mathbb{R}$, and consider at the same time the solution $\phi_\epsilon(x,t)$ of the initial-value problem for the MNLS equation \eqref{eq:MNLS} with wavepacket initial data given by \eqref{eq:wavepacket} and \eqref{eq:Madelungwavepacket}, and the solution $\hat{\phi}_\epsilon(x,t)$ of the initial-value problem for the defocusing NLS equation \eqref{eq:NLS} (with $\kappa=-1$) with wavepacket initial data $\hat{\phi}_0(x)$ given by 
\begin{equation}
\hat{\phi}_0(x)=\hat{A}(x)e^{i\hat{S}(x)/\epsilon},\quad \hat{A}(x):=\sqrt{\rho_0(x)Q(x)},\quad \hat{S}(x):=\hat{S}_0+\int_0^x\left[u_0(y)+2\alpha\rho_0(y)\right]\,dy.
\label{eq:dNLSwavepacket}
\end{equation}
For convenience, we choose the integration constant $\hat{S}_0$ so that $\hat{S}(x)=S(x)+o(1)$ as $x\to -\infty$, that is, we set
\begin{equation}
\hat{S}_0:= S_0 + 2\alpha\int_{-\infty}^0\rho_0(y)\,dy.
\label{eq:hatS0}
\end{equation}

We have already described the scattering data for the MNLS direct spectral problem \eqref{eq:scatODE} with potential $\phi_0$, and now we must do the same for the Zakharov-Shabat spectral problem \eqref{eq:ZSprob} in the self-adjoint case of $\kappa=-1$, with the potential function $\hat{\phi}_0$.
We begin by introducing a shift and $\alpha$-dependent scaling in the Zakharov-Shabat spectral parameter by writing
\begin{equation}
\lambda=\lambda(z):=\frac{2}{\alpha}\left(z-\frac{1}{4}\right).
\label{eq:lambdaz}
\end{equation}
To prepare this spectral problem for WKB analysis we make the transformation 
\begin{equation}
\mathbf{W}^{\mathrm{ZS},-1}_\pm(x;z,\epsilon):=\exp\left(-i\frac{\hat{S}(x)}{2\epsilon}\sigma_3\right)\cdot\mathbf{J}^{\mathrm{ZS},-1}_\pm(x;\lambda(z),\epsilon)
\label{eq:Wmapdefoc}
\end{equation}
analogous to \eqref{eq:Wpmdef} in \eqref{eq:ZSprob} for $\kappa=-1$ (subject to the above substitutions), resulting in the differential equation
\begin{equation}
2\alpha\epsilon\frac{d\mathbf{W}^{\mathrm{ZS},-1}_\pm}{dx}=i\mathbf{M}^{\mathrm{ZS},-1}\mathbf{W}^{\mathrm{ZS},-1}_\pm,
\end{equation}
with coefficient matrix
\begin{equation}
\mathbf{M}^{\mathrm{ZS},-1}:=\begin{bmatrix}-4z+1-\alpha u_0(x)-2\alpha^2\rho_0(x) &-2 i\alpha\sqrt{\rho_0(x)Q(x)}\\-2i\alpha\sqrt{\rho_0(x)Q(x)} &4z-1+\alpha u_0(x)+2\alpha^2\rho_0(x)\end{bmatrix}.
\label{eq:Mdefoc}
\end{equation}
Making the WKB ansatz for column vector solutions of the form $\mathbf{w}=\mathbf{v}(x;z,\epsilon)e^{f(x;z)/(2\alpha\epsilon)}$ leads in the same way as in \S\ref{sec:WKB} to \emph{exactly the same characteristic equation \eqref{eq:characteristic} as governs $f$ in the MNLS case}.
In other words, $\mathbf{M}^{\mathrm{ZS},-1}$ defined by \eqref{eq:Mdefoc} and $\mathbf{M}$
defined by \eqref{eq:Mdef} are similar matrices (in fact they become equal in the limits $x\to\pm\infty$).

By following the line of reasoning described in detail for the MNLS spectral problem in \S\ref{sec:WKB}, one sees easily that the quantity 
\begin{equation}
s^{\mathrm{ZS},-1}(z,\epsilon):=r^{\mathrm{ZS},-1}(\lambda(z),\epsilon),\quad z\in\mathbb{R}
\end{equation}
is small beyond all orders in $\epsilon$ for $z\not\in [z_\mathrm{L},z_\mathrm{R}]$, while
for $z\in (z_\mathrm{L},z_\mathrm{R})$ we have the approximations $s^{\mathrm{ZS},-1}(z,\epsilon)
\sim e^{i\Phi(z)/\epsilon}$ and $1-|s^{\mathrm{ZS},-1}(z,\epsilon)|^2\sim e^{-\tau(z)/\epsilon}$, where $\Phi(z)$ and $\tau(z)$ are given by \emph{exactly the same formulae (\eqref{eq:PhidefWKB} and \eqref{eq:tauWKB}) as in the MNLS case} and where $[z_\mathrm{L},z_\mathrm{R}]$ is the interval in which there exist real turning points.  In other words, at the level of semiclassical approximation, \emph{the MNLS reflection coefficient $s(z;\epsilon)$ is indistinguishable from
the defocusing NLS reflection coefficient $s^{\mathrm{ZS},-1}(z;\epsilon)$}.


For one thing, these considerations show that for supersonic initial data $(\rho_0,u_0)$ used to construct wavepacket initial data for the MNLS equation, the Riemann-Hilbert problem of inverse scattering as described in \S\ref{sec:MNLS-IST} simplifies considerably.  Neglecting the discrete spectrum means that there are no poles in the matrix unknown $\mathbf{N}(z)$.  Moreover, for $z<0$ outside of the interval $[z_\mathrm{L},z_\mathrm{R}]$ the jump condition on the real axis simplifies to $\mathbf{N}_+(z)=i^{\sigma_3}\mathbf{N}_-(z)i^{-\sigma_3}$, while for $z>0$ outside of the interval $[z_\mathrm{L},z_\mathrm{R}]$ the jump condition on the real axis simplifies to the trivial jump $\mathbf{N}_+(z)=\mathbf{N}_-(z)$.  Only in the negative interval $[z_\mathrm{L},z_\mathrm{R}]$ is the jump condition nontrivial, taking the approximate form
\begin{equation}
\mathbf{N}_+(z)=i^{\sigma_3}\mathbf{N}_-(z)i^{-\sigma_3}\begin{bmatrix}
1 & -e^{i(2\theta(z;x,t)+\Phi(z))/\epsilon} \\ e^{-i(2\theta(z;x,t)+\Phi(z))/\epsilon} & e^{-\tau(z)/\epsilon}
\end{bmatrix},\quad z\in (z_\mathrm{L},z_\mathrm{R}).
\label{eq:MNLSdefocjump}
\end{equation}
However, equally impressive is the analogy with the Riemann-Hilbert problem for defocusing NLS with wavepacket initial data of the form \eqref{eq:dNLSwavepacket}.  Indeed, due to the identity $\theta^{\mathrm{ZS}}(\lambda(z);x,t)=\theta(z;x,t)$, the jump condition for the Riemann-Hilbert problem for the latter equation as described in \S\ref{sec:NLS-IST} (and written in terms of the complex variable $z$ rather than $\lambda$) is negligible for $z\not\in [z_\mathrm{L},z_\mathrm{R}]$ while for $z\in (z_\mathrm{L},z_\mathrm{R})$ one has the approximate jump
\begin{equation}
\mathbf{N}^{\mathrm{ZS},-1}_+(\lambda(z))=
\mathbf{N}^{\mathrm{ZS},-1}_-(\lambda(z))\begin{bmatrix}1 & -e^{i(2\theta(z;x,t)+\Phi(z))/\epsilon}\\
e^{-i(2\theta(z;x,t)+\Phi(z))/\epsilon} & e^{-\tau(z)/\epsilon}\end{bmatrix},\quad z\in (z_\mathrm{L},z_\mathrm{R}).
\label{eq:defocjump}
\end{equation}
In both jump conditions \eqref{eq:MNLSdefocjump} and \eqref{eq:defocjump}  there are correction terms we have not written on the off-diagonal that come into play near the endpoints $z_\mathrm{L}$ and $z_\mathrm{R}$ to bring the jump matrix smoothly to the identity; note that $\tau(z_\mathrm{L})=\tau(z_\mathrm{R})=0$ due to coalescence of turning points.  This is a technicality to be handled by the installation of appropriate parametrices in the complex plane near these two points and is not expected to affect the results at leading order in the semiclassical limit.   Setting this technicality aside, we notice that the ``limiting'' Riemann-Hilbert problems for MNLS and defocusing NLS are nearly identical, with the differences being:
\begin{itemize}
\item The point of normalization to the identity for the MNLS Riemann-Hilbert problem is $z=0$, while that for the defocusing NLS Riemann-Hilbert problem is $z=\infty$.
\item The jump condition for the MNLS Riemann-Hilbert problem involves the additional effect of conjugation of $\mathbf{N}_-(z)$ by $i^{\sigma_3}$ for all $z<0$.
\end{itemize}
Regardless of these differences, the same basic asymptotic technique can be applied to both of these problems.  The key technique first appeared in the study of Deift, Venakides, and Zhou \cite{DeiftVZ97} of 
%
%
the small $\epsilon$ behavior of the solution of the Cauchy problem for the Korteweg-de Vries equation
\begin{equation}
\frac{\partial u_\epsilon}{\partial t} + \frac{\partial}{\partial x}\left(\frac{1}{2}u_\epsilon^2\right) = -\epsilon^2\frac{\partial^3 u_\epsilon}{\partial x^3},\quad u_\epsilon(x,0)=u_0(x),
\end{equation}
another kind of problem whose $\epsilon=0$ limit equation is of hyperbolic type, and for which the
associated linear eigenvalue problem in the inverse-spectral transform is self-adjoint.
We first make parallel substitutions of the form $\mathbf{O}(z):=\mathbf{N}(z)e^{ig(z)\sigma_3/\epsilon}$ and $\mathbf{O}^{\mathrm{ZS},-1}(z):=\mathbf{N}^{\mathrm{ZS},-1}(\lambda(z))e^{ig(z)\sigma_3/\epsilon}$ for a suitable and common scalar function $g(z)$ analytic for $z\in\mathbb{C}\setminus [z_\mathrm{L},z_\mathrm{R}]$ and bounded at $z=\infty$.  Since the normalization conditions differ for the MNLS and defocusing NLS problems, we choose (arbitrarily) to preserve the normalization condition in
the MNLS case by imposing the condition that $g(0)=0$.  It is important for semiclassical analysis that $g$ is to be chosen independent of $\epsilon$.  Now, this substitution preserves the domain of analyticity of the unknown matrix functions, and also the MNLS normalization:  $\mathbf{O}(0)=\mathbb{I}$.  The defocusing NLS normalization condition becomes $\mathbf{O}^{\mathrm{ZS},-1}(\infty)=e^{ig(\infty)\sigma_3/\epsilon}$.  The substitution also
preserves the trivial jump conditions $\mathbf{O}_+(z)=\mathbf{O}_-(z)$ for $z>0$, 
$\mathbf{O}_+(z)=i^{\sigma_3}\mathbf{O}_-(z)i^{-\sigma_3}$
for $z<0$ with $z\not\in [z_\mathrm{L},z_\mathrm{R}]$, and $\mathbf{O}^{\mathrm{ZS},-1}_+(z)=\mathbf{O}^{\mathrm{ZS},-1}_-(z)$ for $z\not\in [z_\mathrm{L},z_\mathrm{R}]$.
The jump conditions for $z\in (z_\mathrm{L},z_\mathrm{R})$ take a similar form:
\begin{equation}
\mathbf{O}_+(z)=i^{\sigma_3}\mathbf{O}_-(z)i^{-\sigma_3}\begin{bmatrix} e^{i(g_+(z)-g_-(z))/\epsilon} &
-e^{i(h_+(z)+h_-(z))/\epsilon}\\ e^{-i(h_+(z)+h_-(z))/\epsilon} & e^{i(i\tau(z)-(g_+(z)-g_-(z)))/\epsilon}
\end{bmatrix},\quad z\in (z_\mathrm{L},z_\mathrm{R}),
\label{eq:OMNLSdefocjump}
\end{equation}
and
\begin{equation}
\mathbf{O}^{\mathrm{ZS},-1}_+(z)=\mathbf{O}^{\mathrm{ZS},-1}_-(z)\begin{bmatrix} e^{i(g_+(z)-g_-(z))/\epsilon} &
-e^{i(h_+(z)+h_-(z))/\epsilon}\\ e^{-i(h_+(z)+h_-(z))/\epsilon} & e^{i(i\tau(z)-(g_+(z)-g_-(z)))/\epsilon}
\end{bmatrix},\quad z\in (z_\mathrm{L},z_\mathrm{R}),
\label{eq:Odefocjump}
\end{equation}
where
\begin{equation}
h(z):=\theta(z;x,t)+\frac{1}{2}\Phi(z)-g(z),\quad z\in\mathbb{C}\setminus [z_\mathrm{L},z_\mathrm{R}].
\end{equation}
We insist that $g(z^*)=g(z)^*$, which makes $g_+(z)-g_-(z)$ imaginary and $h_+(z)+h_-(z)$ real.
Now the idea described in \cite{DeiftVZ97} is to try to choose $g(z)$ so that $(z_\mathrm{L},z_\mathrm{R})$ is partitioned into a finite union of three types of subintervals:
\begin{itemize}
\item ``Voids,'' subintervals in which $g_+-g_-=0$ and $h_++h_-=2h$ is strictly increasing.
\item ``Bands,'' subintervals in which $h_++h_-$ is constant and $0<-i(g_+-g_-)<\tau$.
\item ``Saturated regions,'' subintervals in which $g_+-g_-=i\tau$ and $h_++h_-$ is strictly decreasing.
\end{itemize}
Via appropriate factorizations of the jump matrix in voids and saturated regions and subsequent ``steepest descent'' deformations in thin ``lenses'' about these two types of intervals \cite{DeiftVZ97} it can be seen that these intervals do not contribute at leading order to the solution and can be neglected (again, modulo parametrices of Airy type at band endpoints).  In each band subinterval however, the diagonal elements of the jump matrix in \eqref{eq:OMNLSdefocjump} and \eqref{eq:Odefocjump} decay exponentially to zero as $\epsilon\to 0$, while $h_++h_-$ is equal to a fixed real constant (independent of $z$, but generally depending on $x$ and $t$).  Let $h_++h_-=\zeta_n\in\mathbb{R}$ for $z\in (a_n,b_n)\subset (z_\mathrm{L},z_\mathrm{R})$, the
$n^\mathrm{th}$ of $N$ band subintervals, counted from left to right.  This reasoning leads to two related model Riemann-Hilbert problems for the MNLS and defocusing NLS cases.  The model problem
for the MNLS case is to find a $2\times 2$ matrix $\dot{\mathbf{O}}(z)$ with the following properties:
\begin{itemize}
\item[]\textbf{\textit{Analyticity:}} $\dot{\mathbf{O}}(z)$ is analytic for $z\in\mathbb{C}\setminus (-\infty,0]$ and takes boundary values that are continuous except at the band endpoints $\{a_n,b_n\}_{n=0}^{N-1}$ where the matrix elements of $\dot{\mathbf{O}}(z)$ are allowed to blow up like
at worst a negative one-fourth power singularity.
\item[]\textbf{\textit{Jump condition:}}  For $z<0$ outside all bands we have $\dot{\mathbf{O}}_+(z)=i^{\sigma_3}\dot{\mathbf{O}}_-(z)i^{-\sigma_3}$.  In the bands we have instead:
\begin{equation}
\dot{\mathbf{O}}_+(z)=i^{\sigma_3}\dot{\mathbf{O}}_-(z)i^{-\sigma_3}\begin{bmatrix}
0 & -e^{i\zeta_n/\epsilon}\\e^{-i\zeta_n/\epsilon} & 0\end{bmatrix},\quad a_n<z<b_n,\quad n=0,\dots,N-1.
\end{equation}
\item[]\textbf{\textit{Normalization:}} $\dot{\mathbf{O}}(0)=\mathbb{I}$.
\end{itemize}
On the other hand, the model problem for the defocusing NLS case is to find $\dot{\mathbf{O}}^{\mathrm{ZS},-1}(z)$ characterized by the following conditions:
\begin{itemize}
\item[]\textbf{\textit{Analyticity:}} $\dot{\mathbf{O}}^{\mathrm{ZS},-1}(z)$ is analytic for $z\in\mathbb{C}\setminus \cup_{n=0}^{N-1}[a_n,b_n]$ and takes boundary values that are continuous except at the band endpoints $\{a_n,b_n\}_{n=0}^{N-1}$ where the matrix elements of $\dot{\mathbf{O}}^{\mathrm{ZS},-1}(z)$ are allowed to blow up like
at worst a negative one-fourth power singularity.
\item[]\textbf{\textit{Jump condition:}}  
\begin{equation}
\dot{\mathbf{O}}^{\mathrm{ZS},-1}_+(z)=\dot{\mathbf{O}}^{\mathrm{ZS},-1}_-(z)\begin{bmatrix}
0 & -e^{i\zeta_n/\epsilon}\\e^{-i\zeta_n/\epsilon} & 0\end{bmatrix},\quad a_n<z<b_n,\quad n=0,\dots,N-1.
\end{equation}
\item[]\textbf{\textit{Normalization:}} $\dot{\mathbf{O}}^{\mathrm{ZS},-1}(\infty)=e^{ig(\infty)\sigma_3/\epsilon}$.
\end{itemize}
These two model problems are explicitly and uniquely solvable in terms of Riemann theta functions of genus $N-1$.  Let us describe this procedure in detail in the simplest case, $N=1$.
In this case, the solution of the defocusing NLS model problem is completely standard:
\begin{equation}
\dot{\mathbf{O}}^{\mathrm{ZS},-1}(z)=e^{ig(\infty)\sigma_3/\epsilon}e^{i\zeta_0\sigma_3/(2\epsilon)}\begin{bmatrix}1 & 1\\i & -i\end{bmatrix}(z-b_0)^{\sigma_3/4}(z-a_0)^{-\sigma_3/4}
\begin{bmatrix}1 & 1\\i & -i\end{bmatrix}^{-1}e^{-i\zeta_0\sigma_3/(2\epsilon)},
\end{equation}
where the principal branch of the one-fourth power functions is taken in each case.  The solution 
of the MNLS model problem is somewhat nonstandard due to the conjugation by $i^{\sigma_3}$ of the boundary value $\dot{\mathbf{O}}^{\mathrm{ZS},-1}_-(z)$.  It is perhaps easiest to solve this one by reverting from the $z$-plane to the $k$-plane, where $k=z^{1/2}$ (principal branch).  Indeed, if we define
\begin{equation}
\ddot{\mathbf{O}}(k):=\begin{cases}\dot{\mathbf{O}}(k^2),&\quad\Re\{k\}>0\\
i^{\sigma_3}\dot{\mathbf{O}}(k^2)i^{-\sigma_3},&\quad\Re\{k\}<0,
\end{cases}
\end{equation}
then $\ddot{\mathbf{O}}(k)$ has \emph{two} cuts, one on the positive imaginary $k$-axis between the points $i\sqrt{-b_0}$ and $i\sqrt{-a_0}$, and the other on the negative imaginary $k$-axis between $-i\sqrt{-a_0}$ and $-i\sqrt{-b_0}$ (note that $a_0<b_0<0$).  If we denote the boundary value taken on the cuts from the left (respectively right) half-plane as $\ddot{\mathbf{O}}_+(k)$ (respectively $\ddot{\mathbf{O}}_-(k)$), then in both cuts the jump condition turns out to be exactly the same:
\begin{equation}
\ddot{\mathbf{O}}_+(k)=\ddot{\mathbf{O}}_-(k)\begin{bmatrix}0 & e^{i\zeta_0/\epsilon}\\-e^{-i\zeta_0/\epsilon}&0\end{bmatrix},
\end{equation}
and of course we have the normalization condition $\ddot{\mathbf{O}}(0)=\mathbb{I}$.  Solving for $\ddot{\mathbf{O}}(k)$ with jumps of this form is now a standard procedure, and the solution is
\begin{multline}
\ddot{\mathbf{O}}(k)=e^{i\zeta_0\sigma_3/(2\epsilon)}\begin{bmatrix} 1 & 1\\i & -i\end{bmatrix}
(-i(k-i\sqrt{-b_0}))^{\sigma_3/4}(-i(k+i\sqrt{-a_0}))^{\sigma_3/4}\\
{}\cdot(-i(k+i\sqrt{-b_0}))^{-\sigma_3/4}
(-i(k-i\sqrt{-a_0}))^{-\sigma_3/4}\begin{bmatrix}1 & 1\\i & -i\end{bmatrix}^{-1}e^{-i\zeta_0\sigma_3/(2\epsilon)},
\end{multline}
where again the principal branch is intended for all power functions (in particular, this makes the cuts lie on the imaginary axis as intended).  To obtain $\dot{\mathbf{O}}(z)$ we simply restrict $k$ to the right half-plane and write $k=z^{1/2}$ (principal branch).  Thus,
\begin{multline}
\dot{\mathbf{O}}(z)=e^{i\zeta_0\sigma_3/(2\epsilon)}\begin{bmatrix} 1 & 1\\i & -i\end{bmatrix}
(-i(z^{1/2}-i\sqrt{-b_0}))^{\sigma_3/4}(-i(z^{1/2}+i\sqrt{-a_0}))^{\sigma_3/4}\\
{}\cdot(-i(z^{1/2}+i\sqrt{-b_0}))^{-\sigma_3/4}
(-i(z^{1/2}-i\sqrt{-a_0}))^{-\sigma_3/4}\begin{bmatrix}1 & 1\\i & -i\end{bmatrix}^{-1}e^{-i\zeta_0\sigma_3/(2\epsilon)}.
\end{multline}
A completely rigorous semiclassical analysis comes from comparing global parametrices
for $\mathbf{O}(z)$ and $\mathbf{O}^{\mathrm{ZS},-1}(z)$ with the unknown matrices themselves
and proving that the discrepancy in each case is the solution of a ``small norm'' Riemann-Hilbert problem, that is, one with near-identity jump matrices.  In each case, the global parametrix is exactly equal for $|z|$ sufficiently large to the solution of the corresponding model problem we have just presented for $N=1$.  This fact allows us to extract approximate formulae for $\phi_\epsilon(x,t)$ and $\hat{\phi}_\epsilon(x,t)$ that are valid in the semiclassical limit $\epsilon\ll 1$,
by means of the formulae \eqref{eq:MNLSreconstruct} and \eqref{eq:NLSreconstruct}.  It only remains to multiply the model solution on the right by $e^{-ig(z)\sigma_3/\epsilon}$ and calculate asymptotics as $z\to\infty$.  In the defocusing NLS case we have
\begin{equation}
\hat{\phi}_\epsilon(x,t)\sim 2i\lim_{z\to\infty} \lambda(z)\left(\dot{\mathbf{O}}^{\mathrm{ZS},-1}(z)e^{-ig(z)\sigma_3/\epsilon}\right)_{12}=\frac{b_0-a_0}{\alpha}e^{i(2g(\infty)+\zeta_0)/\epsilon},
\end{equation}
from which we obtain approximations to the corresponding Madelung variables in the form:
\begin{equation}
\hat{\rho}_\epsilon(x,t)\sim\hat{\rho}(x,t):=\frac{(b_0-a_0)^2}{\alpha^2}\quad\text{and}\quad
\hat{u}_\epsilon(x,t)\sim\hat{u}(x,t):=\frac{\partial}{\partial x}(2 g(\infty)+\zeta_0).
\label{eq:defocNLSMadelungapprox}
\end{equation}
On the other hand, in the MNLS case we have
\begin{equation}
\phi_\epsilon(x,t)\sim \frac{2}{\alpha}\lim_{z\to\infty}
z^{1/2}\frac{\displaystyle\left(\dot{\mathbf{O}}(z)e^{-ig(z)\sigma_3/\epsilon}\right)_{12}}{\displaystyle\left(\dot{\mathbf{O}}(z)e^{-ig(z)\sigma_3/\epsilon}\right)_{22}}=
\frac{\sqrt{-a_0}-\sqrt{-b_0}}{\alpha}e^{i\zeta_0/\epsilon},
\end{equation}
from which we obtain the Madelung variables
\begin{equation}
\rho_\epsilon(x,t)\sim\rho(x,t):=\frac{(\sqrt{-a_0}-\sqrt{-b_0})^2}{\alpha^2}\quad\text{and}\quad
u_\epsilon(x,t)\sim u(x,t):=\frac{\partial\zeta_0}{\partial x}.
\label{eq:MNLSMadelungapprox}
\end{equation}
Now, the endpoints $(a_0,b_0)$ and the phase constant $\zeta_0$ are certain functions of $x$ and $t$ as required to construct the function $g$.  Indeed, in the case $N=1$ the $z$-derivative of $g$ is necessarily given by the following formula:
\begin{equation}
g'(z)=\frac{S(z)}{2\pi i}\left[\int_{a_0}^{b_0}\frac{\Phi'(s)\,ds}{S_+(s)(s-z)} + \int_{\text{saturated regions}}
\frac{i\tau'(s)\,ds}{S(s)(s-z)}\right]+\theta'(z;x,t)+\frac{8tS(z)}{\alpha^2},
\end{equation}
where $S(z)^2=(z-a_0)(z-b_0)$, $S(z)$ has its branch cut in the interval $[a_0,b_0]$, and $S(z)=z+O(1)$ as $z\to\infty$.  The second integral could be not present at all, or could involve one or both of the intervals $(z_\mathrm{L},a_0)$ and $(b_0,z_\mathrm{R})$.  For arbitrary $(a_0,b_0)$, this formula has a Laurent expansion for large $z$ of the form $g'(z)=M_0(a_0,b_0;x,t) + M_1(a_0,b_0;x,t)z^{-1} +O(z^{-2})$.
The condition that $g'(z)=O(z^{-2})$ as $z\to\infty$ (necessary to ensure that $g(\infty)$ is finite)
then implies that $(a_0,b_0)$ should be determined as functions of $(x,t)$ via the \emph{moment conditions}
\begin{equation}
M_0(a_0,b_0;x,t)=0\quad\text{and}\quad M_1(a_0,b_0;x,t)=0.
\end{equation}
With  the correct configuration of voids and saturated regions, the band endpoints are determined from these conditions as smooth functions of $(x,t)$ by continuation (via the implicit function theorem) from $t=0$, at which time $S(z)^2$ is taken to be the polynomial
\begin{equation}
S(z)^2 = \frac{1}{16}\left[(4z-1+\alpha u_0(x))^2 + 16\alpha^2z\rho_0(x)\right],\quad t=0.
\label{eq:Stzerodefoc}
\end{equation}
(In other words, at $t=0$, the points $a_0(x,0)$ and $b_0(x,0)$ trace out the two branches of the real turning point curve.)  Now with the endpoints determined as functions of $(x,t)$, it is easy
to write down a formula for the $x$-derivative of $g(z)$ (differentiation with respect to $x$ removes the complicated functions $\Phi(z)$ and $\tau(z)$ from the jump conditions):
\begin{equation}
\frac{\partial}{\partial x}g(z) = \frac{\partial\theta}{\partial x}(z;x,t)-\frac{1}{2}\frac{\partial\zeta_0}{\partial x} +\frac{2}{\alpha}S(z).
\end{equation}
Since $g(0)=0$ independently of $x$, we must impose
\begin{equation}
\frac{\partial}{\partial x}(g(0))=0\quad\text{which implies}\quad\frac{\partial\zeta_0}{\partial x}=\frac{1}{\alpha}\left(4\sqrt{-a_0}\sqrt{-b_0}+1\right).
\label{eq:zeta0xeliminate}
\end{equation}
and then it follows easily that
\begin{equation}
\frac{\partial}{\partial x}g(\infty)=\frac{1}{\alpha}\left(\sqrt{-a_0}-\sqrt{-b_0}\right)^2.
\label{eq:gxinfinity}
\end{equation}
The equations \eqref{eq:zeta0xeliminate} and \eqref{eq:gxinfinity} then prove that the semiclassical approximations for $\hat{\rho}_\epsilon(x,t)$ and $\hat{u}_\epsilon(x,t)$ defined
by \eqref{eq:defocNLSMadelungapprox} and those for $\rho_\epsilon(x,t)$ and $u_\epsilon(x,t)$
defined by \eqref{eq:MNLSMadelungapprox} are linked by the identities \eqref{eq:defocmap} for all $(x,t)$ for which the single-band
$g$-function serves to asymptotically and simultaneously reduce both Riemann-Hilbert problems (a condition requiring the auxiliary inequalities on the boundary values of $g$ to be confirmed).
Moreover, the four functions $\hat{\rho}(x,t)$, $\hat{u}(x,t)$, $\rho(x,t)$, and $u(x,t)$ are obviously
explicitly given in terms of $a_0(x,t)$ and $b_0(x,t)$, the latter being determined by a smooth solution to the (algebraic) moment conditions connecting to given supersonic MNLS initial data by the identity \eqref{eq:Stzerodefoc}.  It is an exercise to confirm by differentiation of the moment conditions with respect to $x$ and $t$ that $\rho(x,t)$ and $u(x,t)$ satisfy
the dispersionless MNLS system \eqref{eq:WhithamMNLS}, while $\hat{\rho}(x,t)$ and $\hat{u}(x,t)$ satisfy the dispersionless NLS system \eqref{eq:WhithamNLS} in the defocusing case of $\kappa=-1$.  In other words, the moment conditions provide the solution of the Cauchy initial-value problem for both \eqref{eq:WhithamNLS} and \eqref{eq:WhithamMNLS} in implicit --- but algebraic --- form.

The significance of the analogy between the Riemann-Hilbert problem for MNLS with supersonic initial data and that for defocusing NLS with corresponding wavepacket initial data goes far beyond the case $N=1$ however.  Indeed, the fact that \emph{the very same $g$-function serves to simultaneously reduce both Riemann-Hilbert problems in the semiclassical limit} implies immediately that, for example, when the (essentially common, via the mapping \eqref{eq:defocmap}) solution to the hyperbolic nonlinear dispersionless systems \eqref{eq:WhithamNLS} (for $\kappa=-1$) and \eqref{eq:WhithamMNLS} forms shocks that are
then dispersively regularized by the neglected term $\tfrac{1}{2}\epsilon^2\partial_xF[\rho]$, the oscillations that subsequently develop \emph{will occupy exactly the same space-time region for both the MNLS
and defocusing NLS problems}.  A hint that this could be expected lies in the work of Kuvshinov
and Lakhin \cite{KuvshinovL94} who found a mapping similar to \eqref{eq:defocmap} connecting the (genus one) Whitham modulation equations for MNLS and defocusing NLS.  The latter are derived via an averaging procedure to model four coupled
fields describing the bulk properties of a modulated periodic traveling wave solution, a structure that resembles the wave pattern initially generated by dispersive regularization of shocks.  If one applies the formal matching procedure pioneered by Gurevich and Pitaevskii \cite{GurevichP73} to connect
the Whitham description of a modulated wave with a background described by the dispersionless systems \eqref{eq:WhithamNLS} and \eqref{eq:WhithamMNLS}, one might expect to be able 
to formally establish that the explicit mapping \eqref{eq:defocmap} together with the mapping found by Kuvshinov and Lakhin would give rise to the same space-time region containing at least the first stage of development of the dispersive regularization process.  The power of the observation that the same $g$-function controls both MNLS with supersonic initial data and also
defocusing NLS in the semiclassical limit is that it  makes  the Gurevich-Pitaevskii procedure completely rigorous, and it also shows that similar results hold if shocks subsequently form in the solution of the Whitham modulation equations:  no matter how many breaking curves or caustics appear in the semiclassical dynamics to separate regions modeled by multiphase oscillations of various genera, these caustics will always be exactly the same for both MNLS and defocusing NLS.

\begin{figure}[h]
\begin{center}
\includegraphics{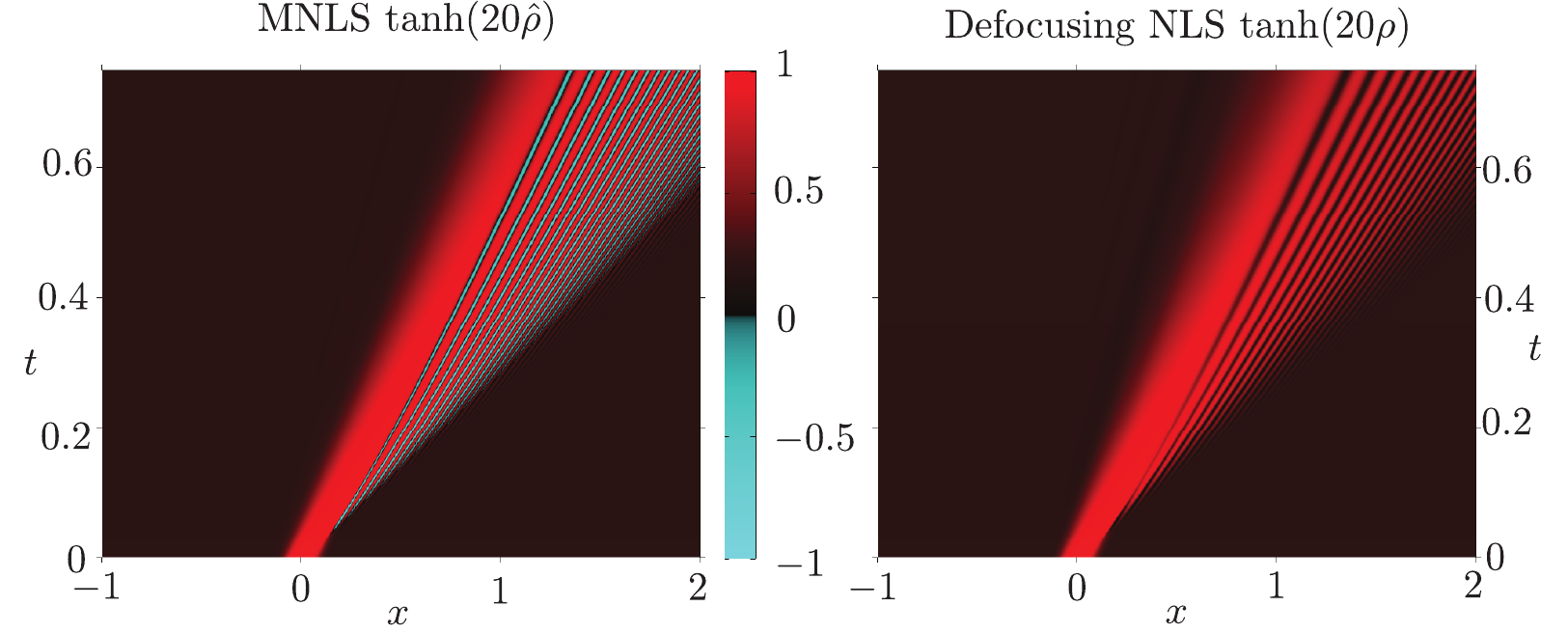}
\end{center}
\caption{Left:  the function $\hat{\rho}_\epsilon(x,t)$ recovered from a simulation of the MNLS equation with supersonic initial data consistent with \eqref{eq:MNLSsupersonicdata}, and with $\alpha=1$ and $\epsilon=0.01$.  Right:  the function $\rho_\epsilon(x,t)$ recovered from a simulation of the defocusing NLS equation for $\epsilon=0.0122$ with initial data obtained from \eqref{eq:MNLSsupersonicdata} via the transformation \eqref{eq:defocmap} with $\alpha=1$.  Dark color indicates small amplitude, while red (green) indicates positive (negative) values.  Note that while as a square modulus $\rho_\epsilon(x,t)$ can never become negative under the defocusing NLS evolution, $\hat{\rho}_\epsilon(x,t)$ can and indeed does become negative under MNLS evolution, although only after wave breaking occurs. Note that the  oscillatory region appears to occupy the same spacetime region in both plots, exactly as predicted by the inverse-spectral theory.  The Matlab code that generated these figures is available from the authors upon request.}
\label{fig:Supersonic}
\end{figure}

The correspondence between the semiclassical MNLS equation with supersonic initial data and the defocusing NLS equation can be illustrated numerically.  We solved the MNLS equation \eqref{eq:MNLS} for $\alpha=1$ with initial data for which the Madelung variables were taken as:
\begin{equation}
\rho_\epsilon(x,0)=\frac{1}{10} + \frac{1}{2}e^{-256x^2}\quad\text{and}\quad u_\epsilon(x,0)=1.
\label{eq:MNLSsupersonicdata}
\end{equation}
Note that since $\alpha=1$ we have $Q_\epsilon(x,0)=\alpha^2\rho_\epsilon(x,0)+\alpha u_\epsilon(x,0)-1=\rho_\epsilon(x,0)>1/10$ so we indeed have a supersonic initial condition.
We took $\epsilon=0.01$, and 
we used a second-order Fourier split-step scheme \cite{Klein08,Tappert74,Yang10} that built in periodic boundary conditions on the domain $x\in [-\pi,\pi]$ with a spatial resolution of $4096$ Fourier harmonics and a time step of $5\cdot 10^{-5}$.  Periodic boundary conditions are numerically convenient and an accurate representation of  fixed boundary conditions if disturbances remain localized.  We chose a small positive initial background value of $1/10$ for $\rho$ because we wanted to be able to recover both $\rho_\epsilon$ and $u_\epsilon$ from the numerical data, and the latter requires careful interpretation if $\phi_\epsilon(x,t)$ becomes small.  The split-step scheme alternately integrates the linear part of the equation exactly in the Fourier domain, and the nonlinear part by the midpoint rule in the physical domain using fixed-point iterations to ensure preservation of the $L^2$-norm and hence maintain accuracy for integration over many time steps as is required to resolve the simulations when $\epsilon\ll 1$ (see \cite{PathriaM90} for a discussion of this important point).  From $\rho_\epsilon(x,t)$ and $u_\epsilon(x,t)$
we calculated $Q_\epsilon(x,t):=\alpha^2\rho_\epsilon(x,t)+\alpha u_\epsilon(x,t)-1$ and then plotted $\hat{\rho}_\epsilon(x,t):=\rho_\epsilon(x,t)Q_\epsilon(x,t)$.  The resulting plot is shown 
in the left-hand panel of Figure~\ref{fig:Supersonic}.
Now according to the correspondence established above with the defocusing NLS equation in the semiclassical limit, we want to compare this simulation with one for the latter equation with initial
data obtained in wavepacket form via the transformation \eqref{eq:defocmap} (with $\alpha=1$ to match the MNLS simulation) of the Madelung initial conditions.  We used virtually the same numerical method and chose $\epsilon=0.0122$ (the slight discrepancy from the MNLS value of $\epsilon$ ensures $2\pi$-periodicity of the initial data for defocusing NLS).  We computed $\rho_\epsilon(x,t)$ by taking the square modulus of the solution and our results are plotted in the right-hand panel of Figure~\ref{fig:Supersonic}.  One sees very clearly that the spacetime region in which the rapid oscillations appear (visible in both plots as a pattern of nearly parallel stripes) is almost the same for both plots.  The theory outlined above indicates that the regions should agree exactly in the semiclassical limit, and also that outside of the (essentially common) oscillation region the plots should agree asymptotically.  The green regions in the left-hand plot of Figure~\ref{fig:Supersonic} are places where $\hat{\rho}_\epsilon(x,t)$ has become negative, which is not inconsistent with our theory as long as it occurs only in the oscillation region for small $\epsilon$.

\section{Subsonic Initial Data}
\label{sec:elliptic}
Now we assume that the initial data for the MNLS equation is globally subsonic for the dispersionless limit system: $Q(x):=\alpha^2\rho_0(x)+\alpha u_0(x)-1<0$ for all $x\in\mathbb{R}$.
Since $\alpha^2\rho_0(x)>0$ for all $x$ and hence in this case $1-\alpha u(x)>0$ for all $x$, it is obvious that the interval $I$ defined in \eqref{eq:Idefine} is a subset of the positive real half-line $\mathbb{R}_+$.  Since in this case the discriminant of the quadratic on the right-hand side
of \eqref{eq:characteristic} is negative for all $x\in\mathbb{R}$, the turning point curve is disjoint from the real axis in the complex $z$-plane.  These facts imply that there are no real turning points $x$ for any real value of $z$, and moreover, the reflection coefficient $s(z,\epsilon)$ is small beyond all orders whenever $z\not\in I$, but is \emph{exponentially large} for $z\in I$.  The
``shadow'' estimate proved in \cite{DiFrancoM08} cannot rule out a significant number of complex eigenvalues in this case, because the shadow of the turning point curve is an $\epsilon$-independent open subset of the complex $z$-plane.  As pointed out earlier, finding both a leading-order estimate for the reflection coefficient in the interval $I$ and good approximations for the discrete spectrum appear to require analyticity of $(\rho_0,u_0)$ as functions of $x$.

This situation should be compared with that arising in the study of semiclassical asymptotics for the initial-value problem for the focusing NLS equation (\eqref{eq:NLS} with $\kappa=+1$) with  wavepacket initial data obtained from the subsonic data $(\rho_0,u_0)$ via the mapping
\eqref{eq:focmap}.  That is, we take initial data for focusing NLS in the form
\begin{equation}
\hat{\phi}_0(x)=\hat{A}(x)e^{i\hat{S}(x)/\epsilon},\quad \hat{A}(x):=\sqrt{-\rho_0(x)Q(x)},\quad
\hat{S}(x):=\hat{S}_0+\int_0^x[u_0(y)+2\alpha\rho_0(y)]\,dy,
\end{equation}
where $\hat{S}_0$ is given by \eqref{eq:hatS0}.  To determine the asymptotic scattering data
for the Zakharov-Shabat problem \eqref{eq:ZSprob} with $\kappa=+1$ corresponding to a wavepacket potential of this form, we make transformations directly analogous to \eqref{eq:lambdaz} and \eqref{eq:Wmapdefoc} to arrive at the equivalent differential equation
\begin{equation}
2\alpha\epsilon\frac{d\mathbf{W}_\pm^{\mathrm{ZS},+1}}{dx}=i\mathbf{M}^{\mathrm{ZS},+1}
\mathbf{W}_\pm^{\mathrm{ZS},+1},
\end{equation}
with coefficient matrix
\begin{equation}
\mathbf{M}^{\mathrm{ZS},+1}:=\begin{bmatrix}-4z+1-\alpha u_0(x)-2\alpha^2\rho_0(x) & 
-2i\alpha\sqrt{-\rho_0(x)Q(x)}\\ 2i\alpha\sqrt{-\rho_0(x)Q(x)} & 4z-1+\alpha u_0(x)+2\alpha^2\rho_0(x)\end{bmatrix}.
\label{eq:Mfoc}
\end{equation}
Seeking column vector solutions $\mathbf{w}=\mathbf{v}(x;z,\epsilon)e^{f(x;z)/(2\alpha\epsilon)}$
and applying the WKB method as in \S\ref{sec:WKB}, one finds that $\mathbf{M}^{\mathrm{ZS},+1}$ defined by \eqref{eq:Mfoc} and $\mathbf{M}$ defined by \eqref{eq:Mdef} are similar matrices and so $f$ satsifies again exactly the same characteristic equation \eqref{eq:characteristic} as in the MNLS case.  The reflection coefficient
\begin{equation}
s^{\mathrm{ZS},+1}(z,\epsilon):=r^{\mathrm{ZS},+1}(\lambda(z),\epsilon),\quad z\in\mathbb{R}
\label{eq:focNLSreflectioncoeff}
\end{equation}
is therefore small beyond all orders for $z\not\in I$ but is large for $z\in I$.  If $\rho_0$ and $u_0$ 
are analytic functions, then it may be possible (depending on the singularity structure of these functions in relation to the location of complex turning points)  to express the asymptotic behavior of the reflection coefficient for $z\in I$ in the form $s^{\mathrm{ZS},+1}(z,\epsilon)\sim e^{i\Phi(z)/\epsilon}$, where $\Phi$ is given by the formula \eqref{eq:PhidefWKB} in which $x_-(z)$ denotes an appropriate complex turning point (this makes $\Re\{i\Phi(z)\}>0$ for $z\in I$).  In these cases the MNLS reflection coefficient $s(z,\epsilon)$ is again indistinguishable from the focusing NLS reflection coefficient \eqref{eq:focNLSreflectioncoeff} in the semiclassical limit.

Let us assume that $I$ is an interval with a nonempty interior.  There are indeed nontrivial cases where this is not so and the reflection coefficient is (essentially) 
everywhere negligible \cite{KamvissisMM03}, but then one necessarily has $O(\epsilon^{-1})$ eigenvalues
that contribute to the inverse-spectral Riemann-Hilbert problem, and the method we are about to explain to remove them from the problem does not apply.  We suppose also that the scattering coefficients $T_{12}(z,\epsilon)$ and $S^{\mathrm{ZS},+1}_{12}(\lambda(z),\epsilon)$ extend analytically from $I$ into a region $R$ of the upper half-plane that contains all of the eigenvalues $\{z_j\}$ for the MNLS problem or $\{\tfrac{1}{2}\alpha\lambda_j+\tfrac{1}{4}\}$ for the defocusing NLS problem.  In this case, the proportionality constants corresponding to the eigenvalues are just given by evaluating $T_{12}(z,\epsilon)$ or $S^{\mathrm{ZS},+1}_{12}(\lambda(z),\epsilon)$
at the eigenvalues.  It follows that upon setting
\begin{equation}
\tilde{\mathbf{N}}^{\mathrm{ZS},+1}(z):=\begin{cases}\displaystyle \mathbf{N}^{\mathrm{ZS},+1}(\lambda(z))\begin{bmatrix}1 & s^{\mathrm{ZS},+1}(z,\epsilon)e^{2i\theta(z;x,t)/\epsilon}\\0 & 1\end{bmatrix},&\quad z\in R\\\\ \displaystyle 
\mathbf{N}^{\mathrm{ZS},+1}(\lambda(z))\begin{bmatrix} 1 & 0\\-s^{\mathrm{ZS},+1}(z^*,\epsilon)^*e^{-2i\theta(z;x,t)/\epsilon} & 1\end{bmatrix},&\quad z\in R^*\\\\
\mathbf{N}^{\mathrm{ZS},+1}(\lambda(z)),&\quad \text{elsewhere},
\end{cases}
\end{equation}
and similarly
\begin{equation}
\tilde{\mathbf{N}}(z):=\begin{cases}\displaystyle \mathbf{N}(z)\begin{bmatrix}1 & s(z,\epsilon)e^{2i\theta(z;x,t)/\epsilon}\\0 & 1\end{bmatrix},&\quad z\in R\\\\ \displaystyle 
\mathbf{N}(z)\begin{bmatrix} 1 & 0\\-s(z^*,\epsilon)^*e^{-2i\theta(z;x,t)/\epsilon} & 1\end{bmatrix},&\quad z\in R^*\\\\
\mathbf{N}(z),&\quad \text{elsewhere},
\end{cases}
\label{eq:MNLSunfold}
\end{equation}
the new matrix unknowns are both pole-free with jumps across the real axis (excluding positive parts of $I$) and the non-real boundary curves $\partial R$ and $\partial R^*$, both of which we take to be oriented from left-to-right, of $R$ and $R^*$ respectively.  Under suitable conditions, the approximate formula $s(z,\epsilon)\sim s^{\mathrm{ZS},+1}(z,\epsilon)\sim e^{i\Phi(z)/\epsilon}$ continues to hold for $z\in \partial R$ (this requires analytic continuation of the complex turning point $x_-(z)$ into the complex $z$-plane)
and we see that the asymptotic jump conditions for $\tilde{\mathbf{N}}(z)$ on $\partial R$ and $\partial R^*$ take the form
\begin{equation}
\tilde{\mathbf{N}}_+(z)=\tilde{\mathbf{N}}_-(z)\begin{bmatrix}1 & -e^{i(2\theta(z;x,t)+\Phi(z))/\epsilon}\\ 0 & 1\end{bmatrix},\quad z\in \partial R
\label{eq:tildeNjump1}
\end{equation}
and
\begin{equation}
\tilde{\mathbf{N}}_+(z)=\tilde{\mathbf{N}}_-(z)\begin{bmatrix}1 &0\\  -e^{-i(2\theta(z;x,t)+\Phi(z))/\epsilon} & 1\end{bmatrix},\quad z\in \partial R^*.
\label{eq:tildeNjump2}
\end{equation}
Here,  the subscript ``$+$'' (respectively ``$-$'') refers to the boundary value taken along the indicated oriented contour from the left (respectively right) side.
Exactly the same asymptotic jump conditions hold on these curves with $\tilde{\mathbf{N}}(z)$ replaced by $\tilde{\mathbf{N}}^{\mathrm{ZS},+1}(z)$.  For the matrix $\tilde{\mathbf{N}}^{\mathrm{ZS},+1}(z)$ there are only asymptotically negligible jumps across the real axis, while for the matrix $\tilde{\mathbf{N}}(z)$ this is true for $z>0$ while for $z<0$ one has
the asymptotic jump condition $\tilde{\mathbf{N}}_+(z)=i^{\sigma_3}\tilde{\mathbf{N}}_-(z)i^{-\sigma_3}$.

To study this problem in the semiclassical limit $\epsilon\ll 1$, we can follow the methodology
of Tovbis, Venakides, and Zhou \cite{TovbisVZ04} by introducing a scalar function $g(z)$, analytic for $z\in\mathbb{C}\setminus (\partial R\cup\partial R^*)$ and independent of $\epsilon$, and then setting, for the same $g$-function, $\mathbf{O}(z):=\tilde{\mathbf{N}}(z)e^{ig(z)\sigma_3/\epsilon}$
and $\mathbf{O}^{\mathrm{ZS},+1}(z):=\tilde{\mathbf{N}}^{\mathrm{ZS},+1}(z)e^{ig(z)\sigma_3/\epsilon}$, which converts the jump conditions \eqref{eq:tildeNjump1}--\eqref{eq:tildeNjump2} into the form
\begin{equation}
\mathbf{O}_+(z)=\mathbf{O}_-(z)\begin{bmatrix}e^{i(g_+(z)-g_-(z))/\epsilon} & -e^{i(h_+(z)+h_-(z))/\epsilon}\\0 & e^{-i(g_+(z)-g_-(z))/\epsilon}\end{bmatrix},\quad z\in\partial R
\end{equation}
and 
\begin{equation}
\mathbf{O}_+(z)=\mathbf{O}_-(z)\begin{bmatrix}e^{i(g_+(z)-g_-(z))/\epsilon} &0\\-e^{-i(h_+(z)+h_-(z))/\epsilon} & e^{-i(g_+(z)-g_-(z))/\epsilon}\end{bmatrix},\quad z\in\partial R^*,
\end{equation}
where $h(z)=\theta(z;x,t)+\tfrac{1}{2}\Phi(z)-g(z)$.
Again, these also hold with $\mathbf{O}(z)$ replaced by $\mathbf{O}^{\mathrm{ZS},+1}(z)$.
The strategy used in \cite{TovbisVZ04} is to choose $g(z)=g(z^*)^*$ so that $\partial R$ splits into
a union of two types of subintervals:
\begin{itemize}
\item ``Voids,'' subintervals in which $g_+-g_-=0$, while $h_++h_-=2h$ has a positive imaginary part.
\item ``Bands,'' subintervals in which $h_++h_-$ is equal to a real constant, while $g_+-g_-$ is
real and strictly increasing along $\partial R$ in the direction of orientation.
\end{itemize}
The jump matrix clearly converges rapidly to the identity matrix as $\epsilon\to 0$ in the union of
voids.  In each band, a three-factor factorization and subsequent ``steepest descent'' lens deformation \cite{TovbisVZ04} reduces the jump matrix to a constant (in $z$) off-diagonal form, with off-diagonal elements of unit modulus.  

We will not go further with the details here, but is should be clear that the use of the very same $g$-function asymptotically reduces both the MNLS and focusing NLS Riemann-Hilbert problems to similar limiting forms $\dot{\mathbf{O}}(z)$ and $\dot{\mathbf{O}}^{\mathrm{ZS},+1}(z)$, respectively whose solutions can be obtained as in \S\ref{sec:hyperbolic}, with the only difference being that the band endpoints now are complex-valued and come in conjugate pairs.  
By estimating the errors carefully as in \cite{TovbisVZ04} we may therefore simultaneously establish the validity for short times independent of $\epsilon$ of the elliptic
dispersionless limit system \eqref{eq:WhithamMNLS} for MNLS with subsonic initial data and also the validity  of the elliptic dispersionless limit system \eqref{eq:WhithamNLS} for  focusing NLS  with corresponding initial data provided via the mapping \eqref{eq:focmap}.  As in the case of supersonic data for the MNLS equation, the subsonic MNLS and focusing NLS equations will display wave breaking phenomena in \emph{exactly the same regions of the $(x,t)$-plane} in the semiclassical limit.

\begin{figure}[h]
\begin{center}
\includegraphics{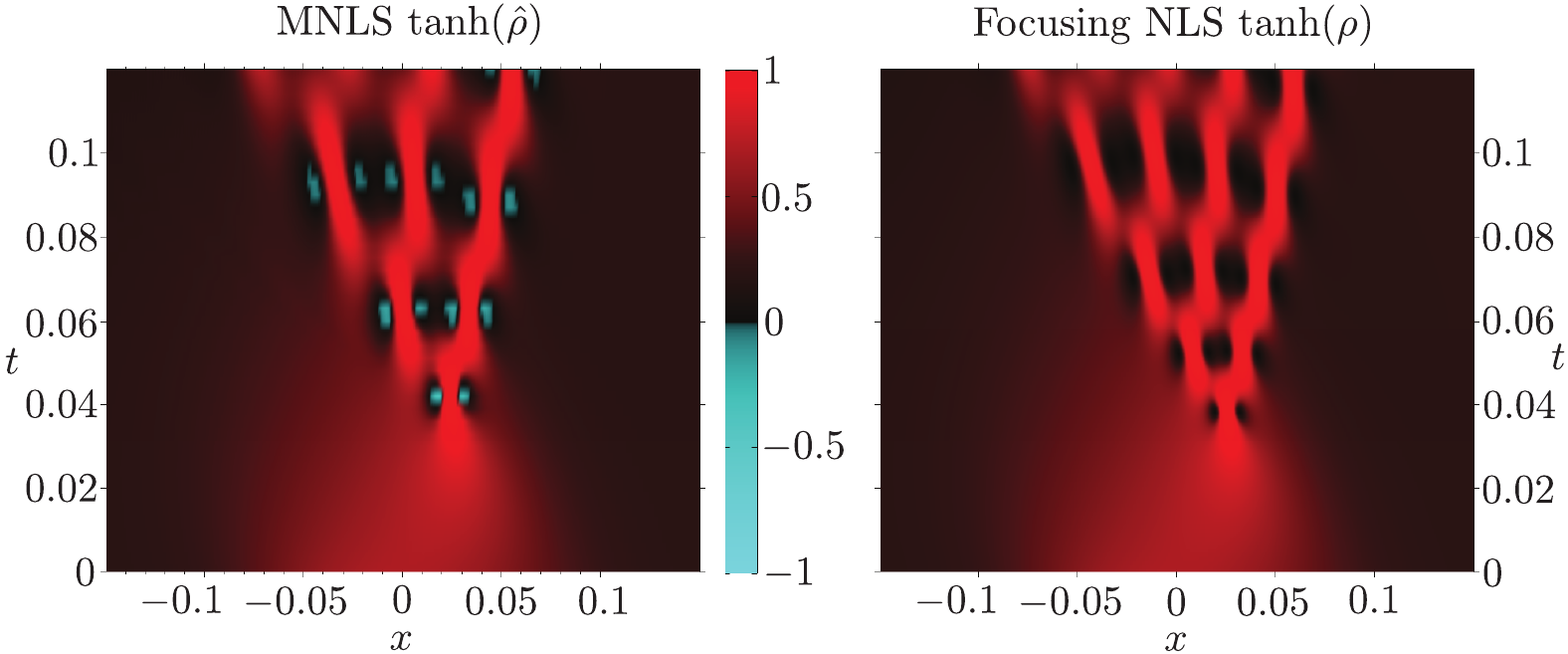}
\end{center}
\caption{Left:  the function $\hat{\rho}_\epsilon(x,t)$ recovered from a simulation of the MNLS equation with subsonic initial data consistent with \eqref{eq:SubsonicIC}, and with $\alpha=1$ and $\epsilon=0.01$.  Right:  the function $\rho_\epsilon(x,t)$ recovered from a simulation of the focusing NLS equation for $\epsilon=0.0078$ with initial data obtained from \eqref{eq:SubsonicIC} via the transformation \eqref{eq:focmap} with $\alpha=1$.  Dark color indicates small amplitude, while red (green) indicates positive (negative) values.  As in Figure~\ref{fig:Supersonic} we see that $\hat{\rho}_\epsilon(x,t)$ can become negative under the MNLS evolution, and also that the oscillations  approximately occupy the same spacetime region in both cases.  The Matlab code that generated these figures is available from the authors upon request.}
\label{fig:Subsonic}
\end{figure}

Again we can illustrate these predictions with numerical simulations.  Consider the initial
data
\begin{equation}
\rho_\epsilon(x,0)=\frac{1}{10} +\frac{1}{2}e^{-256x^2}\quad\text{and}\quad
u_\epsilon(x,0)=-1.
\label{eq:SubsonicIC}
\end{equation}
We take $\alpha=1$ and then it is easy to confirm that $Q_\epsilon(x,0)=\alpha^2\rho_\epsilon(x,0)+\alpha u_\epsilon(x,0)-1\le -7/5<0$ so this indeed corresponds to subsonic wavepacket initial data for the MNLS equation.  The results of a Fourier split-step simulation on $[-\pi,\pi]$ (with $8192$ Fourier modes and a time step of $3\cdot 10^{-5}$) of the MNLS equation with initial data determined up to a constant phase by \eqref{eq:SubsonicIC}
are plotted in the variable $\hat{\rho}_\epsilon(x,t)$ given by \eqref{eq:focmap} in the left-hand 
panel of Figure~\ref{fig:Subsonic}.
Then, we solved the focusing NLS equation (using virtually the same method with $16384$ Fourier modes and a time step of $10^{-5}$) with wavepacket initial data obtained from \eqref{eq:SubsonicIC} via the transformation \eqref{eq:focmap}, and a plot of the square modulus
$\rho_\epsilon(x,t)$ of the solution is shown in the right-hand panel of Figure~\ref{fig:Subsonic}.
The expected correspondence is indeed rather clear.

\section{Transsonic Initial Data}
%
As a specific example of transsonic initial data, consider
\begin{equation}
\rho_0(x):=\mathrm{sech}^2(x)\quad\text{and}\quad u_0(x):=\delta +\mu\tanh(x),
\label{eq:specialdata}
\end{equation}
for suitable real constants $\delta$ and $\mu$.  
It is convenient to introduce in place of $\delta$ and $\mu$ the following normalized parameters:
\begin{equation}
A:=\frac{1-\alpha\delta}{4\alpha^2}\quad\text{and}\quad B:=\frac{\mu}{4\alpha}.
\end{equation}
Let us assume that $A$ and $B$ are subject to the inequalities
\begin{equation}
B^2>A\quad\text{and}\quad B>|A|.
\label{eq:ABineq}
\end{equation}
Then, it is easy to show that for such initial data, $Q(x)=\alpha^2\rho_0(x)+\alpha u_0(x)-1$ is negative for $x<x_\mathrm{c}$
and positive for $x>x_\mathrm{c}$ with a simple root at $x=x_\mathrm{c}=\mathrm{arctanh}(2B-\sqrt{4B^2-4A+1})$.  Therefore, when $t=0$ we are in the subsonic case for $x<x_\mathrm{c}$
and in the supersonic case for $x>x_\mathrm{c}$.  

In \cite{DiFrancoM11}, the Cauchy initial-value problem for the MNLS equation subject to transsonic initial data of this type is carefully studied in the semiclassical limit $\epsilon\ll 1$.
It is easy to obtain specific information about the reflection coefficient $s(z,\epsilon)$ from the WKB
analysis presented in \S\ref{sec:WKB}.  Indeed, only for $z$ in the interval $[z_\mathrm{L},z_\mathrm{R}]$ are there any real turning points at all, where
\begin{equation}
z_\mathrm{L}:=\alpha^2\left(A-\frac{1}{2}-\frac{1}{2}\sqrt{1+4B^2-4A}\right)<z_\mathrm{R}:=0.
\end{equation}
Outside of this interval, $s(z,\epsilon)$ is either exponentially small or exponentially large.
Since $u\pm:=\delta\pm\mu$ in the present situation, according to the discussion around the definition \eqref{eq:Idefine} of the interval $I$, the latter case only occurs for $0<z<\alpha^2(A+B)$.

For both $z_\mathrm{L}<z<0$ and $0<z<\alpha^2(A+B)$ the reflection coefficient is given asymptotically by the formula $s(z,\epsilon)=e^{i\Phi(z)/\epsilon}$ up to a relative error that is
of order $O(\epsilon)$.  In \cite{DiFrancoM11} this approximation is made completely rigorous in a direct fashion; rather than using the WKB method, one can (only for initial data of the special form \eqref{eq:specialdata}) solve the direct spectral problem \eqref{eq:scatODE} exactly in terms of hypergeometric functions \cite{DiFrancoM08}.  This allows the expression of all scattering data for all $\epsilon>0$ in terms of gamma functions \emph{without approximation}, which in turn allows the semiclassical approximation of scattering data to be carried out in full detail with the use of Stirling's formula and its variants.  One of the outcomes of this approach is that there are no eigenvalues whatsoever for any $\epsilon>0$ as a consequence of the first of the two inequalities \eqref{eq:ABineq}. 
The function $\Phi(z)$ defined generally by the WKB formula \eqref{eq:PhidefWKB} turns out to have an analytic continuation from the interval $(z_\mathrm{L},\alpha^2(A+B))$ to the whole upper half $z$-plane.  As a real-valued function in the subinterval $(z_\mathrm{L},z_\mathrm{R})$, this analytic function in turn can be continued to the domain $z\in\mathbb{C}\setminus ((-\infty,z_\mathrm{L}]\cup[z_\mathrm{R},+\infty))$ as a function satisfying the Schwartz-symmetry condition  $\Phi(z^*)=\Phi(z)^*$.  The latter function has the explicit expression (obtained either from \eqref{eq:PhidefWKB} or via Stirling asymptotics):
\begin{equation}
\begin{split}
\Phi(z)&=-4\alpha B\log(2)-\frac{4}{\alpha}(\alpha^2(A+B)-z)\log\left(\frac{2}{\alpha}(\alpha^2(A+B)-z)\right)\\
&\quad\quad{}
+2\left((\alpha^2B^2-z)^{1/2}+\alpha B\right)\log\left(2\left((\alpha^2B^2-z)^{1/2}+\alpha B\right)\right)\\
&\quad\quad{}-2\left((\alpha^2B^2-z)^{1/2}-\alpha B\right)\log\left(2\left((\alpha^2B^2-z)^{1/2}-\alpha B\right)\right)\\
&\quad\quad{}+\left(2(\alpha^2B^2-z)^{1/2}-\frac{2}{\alpha}(z-\alpha^2A)\right)\log\left(2(\alpha^2B^2-z)^{1/2}-\frac{2}{\alpha}(z-\alpha^2A)\right)
\\&\quad\quad{}-\left(2(\alpha^2B^2-z)^{1/2}+\frac{2}{\alpha}(z-\alpha^2A)\right)\log\left(2(\alpha^2B^2-z)^{1/2}+\frac{2}{\alpha}(z-\alpha^2A)\right).
\end{split}
\end{equation}
Here all logarithms and square roots refer to the principal branches. The exponent of the reflection coefficient is obtained from this formula by direct evaluation for $z\in (z_\mathrm{L},z_\mathrm{R}=0)$ and by taking a boundary value from the upper half-plane for $z\in (z_\mathrm{R}=0,\alpha^2(A+B))$.
The latter boundary value satisfies:
\begin{equation}
\Im\{\Phi_+(z)\}=\begin{cases}\displaystyle 2\pi\left((\alpha^2B^2-z)^{1/2}-\alpha B\right),\quad & 0<z<z^+\\\\
\displaystyle\frac{2\pi}{\alpha}(z-\alpha^2(A+B)),\quad &z^+<z<\alpha^2(A+B).
\end{cases}
\end{equation}
Here 
\begin{equation}
z^+:=\alpha^2\left(A-\frac{1}{2}+\frac{1}{2}\sqrt{1+4B^2-4A}\right)\in (0,\alpha^2(A+B)),
\end{equation}
and we note that indeed $\Im\{\Phi_+(z)\}<0$ for $0<z<\alpha^2(A+B)$ as is consistent with 
$s(z,\epsilon)\sim e^{i\Phi_+(z)/\epsilon}$ being exponentially large.

The semiclassical asymptotic analysis of the Riemann-Hilbert problem for MNLS described in \S\ref{sec:MNLS-IST} necessarily combines simultaneously essential elements of both the type of methodology described in \S\ref{sec:hyperbolic} (the case of globally supersonic initial data) and also the type of methodology described in \S\ref{sec:elliptic} (the case of globally subsonic initial data).  One has to factor the jump matrix into two factors in an interval of the form $(w,\alpha^2(A+B))$ 
with $z_\mathrm{L}<w\le z_\mathrm{R}=0$, an interval containing that in which the reflection coefficient is large, and deform the two factors into the upper and lower half-planes
by a transformation like \eqref{eq:MNLSunfold}.  Here one has to be careful because while there are no eigenvalues, there exist poles of the scattering coefficient $T_{12}(z,\epsilon)$ emanating
from the point $z=\alpha^2(A+B)$ into the upper half-plane, and the transformation \eqref{eq:MNLSunfold} will \emph{introduce} artificial poles (in \cite{DiFrancoM08} they are called ``phantom poles'', but they are also similar to ``false poles'' \cite{ReedS79III} or ``resonance poles'' \cite{ReedS79IV} in scattering theory).  A similar situation occurs in the semiclassical analysis of focusing NLS in \cite{TovbisVZ04}, and the authors of that work argue that the region $R$ can be chosen to abut the line of poles only at the right endpoint of the interval in which the reflection coefficient is large; this reduces the problem to a local one near the endpoint, and there an implicit parametrix is constructed to deal with the breakdown of the semiclassical approximation of the reflection coefficient.  By contrast, in \cite{DiFrancoM11} a local construction is used that actually allows poles to reside in $R$ (by a modified pole-removing substitution) and controls all errors with exponential accuracy near $z=\alpha^2(A+B)$ without the use of any local parametrix whatsoever.

Once the large reflection coefficient has been removed from the real axis in this way, one introduces a $g$-function that now has the job of controlling \emph{at the same time} the triangular jump matrix factors on $\partial R$ and $\partial R^*$ as in \S\ref{sec:elliptic} and the oscillatory full jump matrix in the real interval $(z_\mathrm{L},z_\mathrm{R}=0)$ as in \S\ref{sec:hyperbolic}.  
One of the results of \cite{DiFrancoM11} is that the transsonic point $x_\mathrm{c}$ actually persists for
small nonzero $t$ as a moving boundary $x=x_\mathrm{c}(t)$ that separates asymptotically
subsonic flow for $x<x_\mathrm{c}(t)$ from asymptotically supersonic flow for $x>x_\mathrm{c}(t)$.
In the gas dynamics literature, such a moving boundary is called a \emph{sonic line}.
At the level of the Riemann-Hilbert problem and its $g$-function, what happens as one crosses the sonic line from left-to-right is that the complex endpoint $q=q(x,t)$ of a single band on $\partial R$ connecting $z=q$ to $z=\alpha^2(A+B)$ collides with its complex conjugate at a point $z_\mathrm{c}(t)\in (z_\mathrm{L},z_\mathrm{R})$ and the pair re-opens as a real band $(a_0(x,t),b_0(x,t))$ surrounded by voids; the point $w$ at which $\partial R$ meets the real axis in the interval $(z_\mathrm{L},z_\mathrm{R})$ is then taken as the right endpoint $z=b_0$ of the newly-born real band.
To determine the sonic line, one first solves the equation
\begin{equation}
\frac{16t}{\alpha^2}-\Phi''(z_\mathrm{c}) +\frac{2}{\pi}\int_0^{\alpha^2(A+B)}\frac{\Im\{\Phi_+'(s)\}\,ds}{(s-z_\mathrm{c})^2}=0
\label{eq:zcsonicline}
\end{equation}
for $z_\mathrm{c}=z_\mathrm{c}(t)$ (for $t=0$ this gives $z_\mathrm{c}=\alpha^2(2B^2-A-\sqrt{4B^2-4A+1})\in (z_\mathrm{L},0)$, which can be used as a starting point for Newton iteration to obtain the solution for small nonzero $t$), and then the sonic line $x=x_\mathrm{c}(t)$ is given explicitly by
\begin{equation}
x_\mathrm{c}(t)=\frac{1}{\alpha}\left(1-4z_\mathrm{c}(t)\right)t+\frac{\alpha}{4}\Phi'(z_\mathrm{c}(t)) -\frac{\alpha}{2\pi}\int_0^{\alpha^2(A+B)}\frac{\Im\{\Phi_+'(s)\}\,ds}{s-z_\mathrm{c}(t)}.
\label{eq:xcsonicline}
\end{equation}

In the vicinity of the sonic line, and in fact throughout the subsonic flow region to its left, the $\epsilon$-dependent functions $\rho_\epsilon(x,t)$ and $u_\epsilon(x,t)$ are accurately approximated in the limit $\epsilon\ll 1$ by a certain solution $\rho(x,t)$ and $u(x,t)$ of the mixed-type dispersionless MNLS system \eqref{eq:WhithamMNLS}.
Moreover, the solution is characterized without dealing with partial differential equations at all by
the following formulae.  For arbitrary $\rho>0$ and $u\in\mathbb{R}$, let $S(z;\rho,u)$ be the function determined by the conditions that
\begin{equation}
16S(z;\rho,u)^2=(4z-1+\alpha u)^2 + 16\alpha^2\rho z
\label{eq:quadratic}
\end{equation}
and that $S(z;\rho,u)$ is analytic in the complex plane excluding a branch cut along the straight line connecting the two roots of the quadratic \eqref{eq:quadratic}, with asymptotic behavior
$S(z;\rho,u)=z+O(1)$ as $z\to\infty$.  Then define
\begin{equation}
\begin{split}
M_0(\rho,u;x,t)&:=\frac{2\pi}{\alpha^2}(\alpha x - (\alpha u+2\alpha^2\rho)t)-
\frac{1}{4i}\oint\frac{\Phi'(s)\,ds}{S(s;\rho,u)} +\int_0^{\alpha^2(A+B)}\frac{\Im\{\Phi_+'(s)\}\,ds}{S(s;\rho,u)}\\
M_1(\rho,u;x,t)&:=\frac{\pi}{2\alpha^2}\left((1-\alpha u-2\alpha^2\rho)\alpha x+\left(6\alpha^4\rho^2+6\alpha^3\rho u-4\alpha^2\rho+\alpha^2u^2-\alpha u\right)t\right)\\
&\quad\quad{} - 
\frac{1}{4i}\oint\frac{\Phi'(s)s\,ds}{S(s;\rho,u)} + \int_0^{\alpha^2(A+B)}\frac{\Im\{\Phi_+'(s)\}s\,ds}{S(s;\rho,u)}.
\end{split}
\label{eq:Moments}
\end{equation}
Here, the contour integrals are taken over a positively-oriented loop surrounding the branch cut of $S$.  One then solves the equations $M_0=M_1=0$ by Newton iteration for $(\rho,u)$ as functions of $(x,t)$ near $t=0$ 
with the initial conditions $\rho(x,0)=\rho_0(x)$ and $u(x,0)=u_0(x)$.  The Jacobian of this system of equations vanishes along the sonic line, even though the solution $\rho(x,t)$ and $u(x,t)$ remains smooth.  This means that Newton iteration to find $(\rho,u)$ will fail for
$x=x_\mathrm{c}(t)$, and will be numerically difficult (will require a very accurate initial guess for convergence) for $(x,t)$ close to the sonic line $x=x_\mathrm{c}(t)$.  However, along the sonic line exactly, $\rho(x_\mathrm{c}(t),t)$ and $u(x_\mathrm{c}(t),t)$ can be found directly from 
the auxiliary function $z_\mathrm{c}(t)$ solving \eqref{eq:zcsonicline}:
\begin{equation}
\rho(x_\mathrm{c}(t),t)=-\frac{4}{\alpha^2}z_\mathrm{c}(t)\quad\text{and}\quad
u(x_\mathrm{c}(t),t)=\frac{1}{\alpha}(1+4z_\mathrm{c}(t)).
\label{eq:rho0u0sonicline}
\end{equation}
It is easy to check that upon elimination of $z_\mathrm{c}$ between these two equations one arrives at the condition $Q=\alpha^2\rho+\alpha u-1=0$, the condition for degeneration of the type of the quasilinear system \eqref{eq:WhithamMNLS}.


\begin{thebibliography}{99}
  \bibitem{DeiftVZ97} P. Deift, S. Venakides, and X. Zhou, ``New results in small dispersion KdV by an extension of the steepest descent method for Riemann-Hilbert problems,'' \textit{Internat. Math. Res. Notices}, \textbf{1997}, 286--299, 1997.
\bibitem{DiFrancoM08} J. C. DiFranco and P. D. Miller, ``The semiclassical modified nonlinear Schr\"odinger equation I:  Modulation theory and spectral analysis,'' \textit{Physica D}, \textbf{237}, 947--997, 2008.
\bibitem{DiFrancoM11} J. C. DiFranco and P. D. Miller, ``The semiclassical modified nonlinear Schr\"odinger equation II:  Asymptotic analysis of the Cauchy problem,'' in preparation, 2011.
\bibitem{GurevichP73} A. V. Gurevich and L. P. Pitaevskii, ``Nonstationary structure of a collisionless shock wave,'' \textit{Sov. Phys. JETP}, \textbf{38}, 291--297,  1974. 
\bibitem{KamvissisMM03} S. Kamvissis, K. T.-R. McLaughlin, and P. D. Miller,
  \textit{Semiclassical Soliton Ensembles for the Focusing Nonlinear
    Schr\"odinger Equation}, Annals of Mathematics Studies
  \textbf{154}, Princeton University Press, Princeton, 2003.
  \bibitem{Klein08} C. Klein, ``Fourth order time-stepping for low dispersion Korteweg-de Vries and nonlinear Schr\"odinger equations,'' \textit{Electron. Trans. Numer. Anal.}, \textbf{29}, 116--135, 2008.
\bibitem{KuvshinovL94}  B. N. Kuvshinov and V. P. Lakhin, ``The Riemann invariants and characteristic velocities of Whitham equations for the derivative nonlinear Schr\"odinger equation,'' 
\textit{Phys. Scr.},  \textbf{49}, 257--260,  1994. 
\bibitem{Lax86}
P. D. Lax, ``On dispersive difference schemes. Solitons and coherent structures'' (Santa Barbara, 
Calif., 1985). \textit{Physica D}, \textbf{18}, 250--254, 1986. 
\bibitem{LaxL83}P. D. Lax and C. D. Levermore, ``The small dispersion limit of the Korteweg-de Vries equation,''  \textit{Comm. Pure Appl. Math.}, \textbf{36}, 253--290 (Part I), 571--593 (Part II), 809--929 (Part III), 1983. 
\bibitem{Madelung26}
E. Madelung, ``Quantum theory in hydrodynamic form,'' \textit{Zeitschrift f\"ur Physik}, \textbf{40}, 322--326, 1926.
\bibitem{Miller01} P. D. Miller, ``Some remarks on a WKB method for the nonselfadjoint Zakharov-Shabat eigenvalue problem with analytic potentials and fast phase,'' \textit{Physica D}, \textbf{152--153}, 145--162, 2001.
\bibitem{PathriaM90} D. Pathria and J. Morris, ``Pseudo-spectral solution of nonlinear Schr\"odinger equations,'' \textit{J. Comp. Phys.}, \textbf{87}, 108--125, 1990.
\bibitem{ReedS79III} M. Reed and B. Simon, \textit{Methods of Modern Mathematical Physics, Volume III:  Scattering Theory}, Academic Press, San Diego, CA, 1979.
\bibitem{ReedS79IV} M. Reed and B. Simon, \textit{Methods of Modern Mathematical Physics,
Volume IV:  Analysis of Operators}, Academic Press, San Diego, CA, 1979.
\bibitem{Tappert74} F. Tappert, ``Numerical solutions of the Korteweg-de Vries equation and its generalizations by the split step Fourier method,'' in \textit{Nonlinear Wave Motion}, AMS Lectures in Applied Mathematics, Volume 15 (A. C. Newell, ed.), 215--216, 1974.
\bibitem{TovbisVZ04}
A. Tovbis, S. Venakides, and X. Zhou, 
``On semiclassical (zero dispersion limit) solutions of the focusing nonlinear Schr\"odinger equation,''
\textit{Comm. Pure Appl. Math.},  \textbf{57}, 877--985, 2004. 
\bibitem{Yang10} J. Yang, \textit{Nonlinear Waves in Integrable and Nonintegrable Systems}, 
Mathematical Modeling and Computation, Volume 16, Society for Industrial and Applied Mathematics, Philadelphia, 2010.
\bibitem{ZakharovS72} V. E. Zakharov and A. B. Shabat, ``Exact theory of two-dimensional self-focusing and one-dimensional self-modulation of waves in nonlinear media,'' \textit{Sov. Phys. JETP}, \textbf{34}, 62--69, 1972.
\end{thebibliography}
\end{document}